\documentclass[a4paper,11pt]{article}
\usepackage{graphicx}
\usepackage{xcolor}
\usepackage{amsmath,amsfonts,amssymb,amstext,graphicx}
\usepackage[hidelinks]{hyperref}
\usepackage{cancel}
\usepackage{empheq}
\usepackage{float}
\usepackage{cite}
\usepackage{subcaption}
\usepackage{authblk}
\usepackage{comment}
\usepackage{blindtext}
\usepackage{cases}
\usepackage{ulem}
\usepackage{epsfig}%
\usepackage[bottom]{footmisc}
\usepackage{lineno}

\usepackage{epsfig}%
\usepackage[bottom]{footmisc}
\begin{document}
\title{\centerline \textbf {Observational constraints on generic models of non-minimal curvature-matter coupling}}

\author[1]{ Anirban Chatterjee\thanks{ Corresponding author: iitkanirbanc@gmail.com, anirbanchatterjee@nbu.edu.cn}}
\author[2]{Akshay Panda\thanks{akshay.panda@gm.rkmvu.ac.in}}
\author[2]{Abhijit Bandyopadhyay\thanks{abhijit.phy@gm.rkmvu.ac.in}}
\affil[1]{Institute of Fundamental Physics and Quantum Technology,\par
 Department of Physics, School of Physical Science and Technology,\par 
  Ningbo University, Ningbo, Zhejiang 315211, China}\par

\affil[2]{Department of Physics, Ramakrishna Mission Vivekananda Educational and Research Institute, Belur Math, Howrah-711202, West-Bengal, India}

\date{\today}
\maketitle

\begin{abstract}
We investigate two classes of non-minimally coupled curvature-matter models in the FLRW universe with a perfect fluid and analyze their cosmological implications using Supernova Ia, Observed Hubble Data, and Baryon Acoustic Oscillation measurements. Non-minimal coupling is introduced via an additional term
$\int d^4x \sqrt{-g} \mathcal{G}({\cal L}_{m}) f_2(R)$ 
in the Einstein-Hilbert action.
To obtain observational constraints, we use an exponential-type fluid-pressure profile $p = p_0e^{ak}$
characterized by the dimensionless parameter $k$ 
 and parameterize $f_2(R)$ as $R^n$
with another dimensionless parameter $n$.
Two additional parameters,  $\alpha$ and $\beta$
in the functional form of 
$\mathcal{G}({\cal L}_{m})$ determine the coupling strength.
We identify significant regions in the $(n, k)$-parameter space
for fixed coupling strength values where non-minimally coupled models align with observed late-time cosmic evolution. Additionally, we explore and discuss features of energy transfer   between the curvature and matter sectors using observational data. 
\end{abstract}

%\linenumbers

\section{Introduction}  
\label{Sec:I}

  Observations of Type Ia Supernovae (SNe Ia) by Riess et al. (1998) \cite{ref:Riess98} and Perlmutter et al. \cite{ref:Perlmutter} confirm that the universe is currently undergoing accelerated expansion, with a transition from deceleration to acceleration occurring in the late-time cosmic evolution. This acceleration is attributed to ``Dark Energy" (DE), an unclustered form of energy driving late-time acceleration. 
Evidence for ``Dark Matter" (DM), a non-luminous matter detected via gravitational interactions, comes from spiral galaxy rotation curves \cite{Sofue:2000jx}, gravitational lensing \cite{Bartelmann:1999yn}, and colliding clusters like the Bullet Cluster \cite{Clowe:2003tk}. Experiments such as WMAP \cite{Hinshaw:2008kr} and Planck \cite{Ade:2013zuv} show that DE and DM contribute approximately 96\% of the total energy density of the universe, with 69\% from DE, 27\% from DM, and the remaining 4\% from radiation and baryonic matter. 
 Einstein introduced the cosmological constant term $\Lambda g_{\mu\nu}$ into his field equations to model a static universe but later abandoned it following Hubble's observation of an expanding universe. The discovery of late-time cosmic acceleration renewed interest in $\Lambda$, as it provides a straightforward solution for accelerated expansion. The corresponding phenomenological model, the $\Lambda$-CDM model (where `CDM' denotes cold dark matter), effectively describes cosmic acceleration.
However, the $\Lambda$-CDM model faces issues such as the coincidence problem \cite{Zlatev:1998tr} and the fine-tuning problem \cite{Martin:2012bt}, motivating the exploration of alternative explanations for dark energy.\\

Dark energy models can be broadly categorized into two main types. The first involves field-theoretic modifications to the energy-momentum tensor in Einstein's equations, introducing a field as an additional component along with matter and radiation. These include quintessence \cite{Peccei:1987mm,Ford:1987de,Peebles:2002gy,Nishioka:1992sg,Ferreira:1997au,Ferreira:1997hj,Caldwell:1997ii,Carroll:1998zi,Copeland:1997et,Hussain:2023kwk,Chatterjee:2024duy} and $k$-essence models \cite{Fang:2014qga,ArmendarizPicon:1999rj,ArmendarizPicon:2000ah,ArmendarizPicon:2000dh,ArmendarizPicon:2005nz,Chiba:1999ka,ArkaniHamed:2003uy,Caldwell:1999ew,Bandyopadhyay:2017igc,Bandyopadhyay:2018zlz,Bandyopadhyay:2019ukl,Bandyopadhyay:2019vdd,Chatterjee:2021hhj,Chatterjee:2022uyw,Chatterjee:2021ijw,Hussain:2022osn,Bhattacharya:2022wzu}. 
The second type alters the geometric part of Einstein's equations, modifying the Einstein-Hilbert action. These include $f(R)$ gravity models \cite{fr1,fr2,fr3,fr4,fr5,fr6,fr7,fr8,fr9}, scalar-tensor theories, Gauss-Bonnet gravity, and braneworld models. Most scenarios assume a homogeneous and isotropic universe described by the Friedmann-Lema\^itre-Robertson-Walker (FLRW) metric with a scale factor $a(t)$ and curvature constant $K$, though some consider an inhomogeneous universe with a perturbed FLRW metric.\\

In this article, we explore a model of the universe with a perfect fluid content that is non-minimally coupled to spacetime curvature. 
The perfect fluid is characterized by its energy density $(\rho)$ and pressure $(p)$.
We introduce a non-minimal curvature-matter coupling by adding the term 
$\int d^4x \sqrt{-g} \mathcal{G}({\cal L}_{m}) f_2(R)$
to the action, which includes both the Einstein-Hilbert action and the minimally coupled matter action.   Here, $R$ 
is the Ricci Scalar  and $\mathcal{G}({\cal L}_{m})$ is  
a function of the matter Lagrangian of the hypothetical fluid. 
By selecting an appropriate functional form for the matter Lagrangian, we can establish a connection between the curvature and matter sectors. We investigate the constraints on the parameters involved in this type of curvature-fluid coupled model using a combined analysis of the Pantheon compilation of 1048 SNe Ia data points \cite{Pan-STARRS1:2017jku}, 54 data points from Observed Hubble data \cite{Geng:2018pxk}, and BAO datasets \cite{WiggleZ}.\\

In the broader scope of modified gravity $f(R)$-based models, the geometric component of Einstein's equation is modified by altering the Einstein-Hilbert action to $\int d^4x \sqrt{-g} f(R)$, where $R$ is replaced by a function $f(R)$ \cite{Harko:2008qz, Bertolami:2008ab,Schutz:1970,Brown:1993}. In the realm of $f(R)$ gravity theories, it has been demonstrated \cite{Bertolami:2007} that the covariant derivative of the energy-momentum tensor does not vanish ($\nabla_{\mu}T^{\mu \nu} \neq 0$), particularly when there is a curvature-matter coupling, potentially leading to a deviation from geodesic motion due to the emergence of a new force. The implications of such models on stellar equilibrium have been explored in \cite{Bertolami:2007vu, Sotiriou:2008dh}. An analogy has been established between an appropriate scalar theory and the typical model featuring a non-minimal curvature-matter coupling \cite{Bertolami:2008im}. In \cite{Faraoni:2007sn}, criteria to prevent instabilities in such models have been thoroughly discussed. The dynamics of particles and fields influenced by curvature-matter couplings have been analyzed in \cite{Sotiriou:2008it}.
A natural logical choice for the matter Lagrangian, as suggested in \cite{Bertolami:2007vu, Sotiriou:2008dh}, is of the form ${\cal L}_m = p$, where $p$ represents the pressure of the matter fluid. This choice effectively replicates the correct hydrodynamic equations of a perfect fluid and, within the framework of curvature-matter coupling, leads to the elimination of any additional forces. However, alternative choices that may also lead to the elimination of this extra force are possible, as discussed in \cite{Brown:1993, Hawking:1973}. 
In this study, we adopted  a generic form of the function $\mathcal{G}(\mathcal{L}_m)$ in the interaction, which is coupled multiplicatively with the modified $f_2(R)$ function. This form of $\mathcal{G}(\mathcal{L}_m)$ was deliberately chosen to ensure that $\frac{\partial \mathcal{G}}{\partial \mathcal{L}_m} \mathcal{L}_m \neq \mathcal{G}$. To satisfy this non-equality condition, we adopted two distinct forms of $\mathcal{G}(\mathcal{L}_m)$. The linear form results in a difference $\frac{\partial \mathcal{G}}{\partial \mathcal{L}_m} \mathcal{L}_m - \mathcal{G}$ that is a non-zero constant ($\beta$), while the non-linear form of $\mathcal{G}(\mathcal{L}_m)$ yields $\frac{\partial \mathcal{G}}{\partial \mathcal{L}_m} \mathcal{L}_m - \mathcal{G}$ to be proportional to $\mathcal{L}_m$, with a proportionality constant $\alpha$ serving as a coupling parameter for the non-linear scenario.  We may mention here that, in contemporary studies \cite{Bertolami:2008ab}, \cite{Bertolami:2007vu}, \cite{Bertolami:2008im, Faraoni:2007sn, Sotiriou:2008it},\cite{Bertolami:2007gv, Dolgov:2003fw} modification of $f(R)$ gravity theories was introduced by coupling an arbitrary function of the Ricci scalar $R$ with the matter Lagrangian density ${\cal L}_m$. This results in non-geodesic motion for massive particles and an additional force orthogonal to the four-velocity. The connection to MOND and the Pioneer anomaly was explored. The effects of non-minimal coupling on stellar equilibrium were studied, with constraints on the coupling derived. An inequality to prevent the Dolgov-Kawasaki instability was established. The relationship between geometry-matter coupling models and scalar-tensor theories was analyzed, including implications for dark matter. When the action and coupling are linear in $R$, the theory introduces higher-order derivatives of matter fields without affecting gravity. The equivalence with scalar theories allowed calculation of the post-Newtonian parameters $\beta$ and $\gamma$. Various forms of ${\cal L}_m$ and their associated extra-forces were examined, showing that natural forms of ${\cal L}_m$ do not eliminate the extra-force. These couplings have also been explored as explanations for dark energy and the universe's accelerated expansion. In this work we considered  the matter Lagrangian density as ${\cal L}_m = p$ and defining $f_2(R) = R^n$, we  derived relationships between $a(t)$, $\rho(t)$, and $p(t)$ using their time derivatives, along with other quantities such as the Hubble function $H(t)$ and the Ricci scalar $R(t)$. These equations are utilized to explore constraints on non-minimal models through the analysis of observational data.
\\

We derived a model-independent parametrization of the Hubble parameter $H$ in terms of redshift $z$ by analyzing various observational datasets (Pantheon, OHD, BAO). By normalizing the FLRW scale factor $a$ to 1 at the present epoch ($z=0$) using the relation $\frac{1}{a} = 1+z$, we can express the temporal behavior of cosmological quantities in terms of $z$, $a$, or a dimensionless time parameter $\tau \equiv \ln a$.  By analyzing the Hubble parameter profile over the redshift range relevant to the SNe Ia data, we determined the time evolution of the FLRW scale factor $a$ and its higher-order time derivatives $\dot{a}$ and $\ddot{a}$. This approach enabled us to examine the temporal evolution of the Ricci scalar $R$ and its time derivatives, which are crucial for our analysis. \\

To investigate observational constraints on a non-minimally coupled curvature-fluid model, characterized by the parameters $n$ from the curvature sector and $k$ from the matter sector, while keeping the coupling parameters ($\alpha$, $\beta$) constant, we considered a fluid pressure model where the temporal behavior of pressure $p$ is modeled as $p \sim e^{ak}$, with $k$ being a dimensionless parameter. Consequently, the evolution equations were influenced by these four parameters ($\alpha$, $\beta$, $n$, $k$), affecting the model-based computations of $\rho$ and $p$.
Using observational inputs, we determined regions in terms of parameters from the curvature and matter sectors where the coupled system satisfies different classical energy conditions. Depending on these conditions, we identified three distinct regions in the parameter space. From these regions, we selected several benchmark points and plotted the effective equation of state (EoS) parameter of the coupled system. We also assessed how this coupled system deviates from the original $\Lambda$-CDM model  in the context of the usual minimal coupling scenario. 
The effective equation of state (EoS) parameter for the coupled sector is defined by the ratio of pressure to energy density. We examined the temporal variation of the  effective EoS parameter using specific benchmark values of the model parameters ($n, k$), selected from permissible regions of parameter space ensuring that the modified energy density parameter remains positive throughout all observable instances. The classification of the parameter space into three regions is based on classical energy conditions: the first region ($C_1$) includes cases where $\rho \geq 0$ (which satisfies Weak Energy Condition (WEC)), the second region ($C_2$) covers conditions where $\rho \geq 0$, $P < 0$, and $\rho + P \geq 0$ (which includes  Dominant Energy Condition (DEC), Null Energy Condition (NEC) and Weak Energy Condition (WEC)), and the third region ($C_3$) represents scenarios where $\rho \geq 0$, $\rho + P \geq 0$, and $\rho + 3P \leq 0$ (which includes Null Energy Condition (NEC) and Weak Energy Condition (WEC) but violates Strong Energy Condition (SEC)). For both coupling types, we selected three benchmark values corresponding to the parameters $k$ and $n$ from regions $C_2$ and $C_3$, and analyzed the temporal behavior of  effective EoS parameter using a logarithmic function of the FLRW scale factor.  In this curvature-matter coupled scenario, the curvature sector, governed by \( f_2(R) = R^n \), is connected to the matter sector, modeled with an exponential fluid pressure parameter \( k \). Coupling parameters \( \alpha \) and \( \beta \) drive the dynamics of both sectors. From the modified Friedmann equation, energy densities and pressure are expressed in terms of \( n \), \( k \), \( \alpha \), and \( \beta \). Using observational data, the model is constrained for both linear and non-linear coupling regimes, with benchmark values of \( \alpha \) and \( \beta \) aiding in parameter determination.  Analysis reveals that the model primarily depicts non-phantom dark energy behavior in the effective equation of state (EoS), consistent with the observational redshift range.  \\

In the context of non-minimally coupled curvature-matter scenarios, the presence of a non-zero covariant derivative of the energy-momentum tensor indicates an exchange of energy between the curvature and matter domains. We computed the energy exchange rate for two types of coupling cases using specific benchmark values of the model parameters ($\alpha, \beta, n, k$). The relatively higher magnitude of the energy exchange rate during the early phase compared to that of the late-time phase suggests that the interaction between curvature and matter dominates in the early stages of evolution. Previous studies have explored the thermodynamic implications of curvature-matter coupling scenarios in cosmology \cite{Harko:2015pma, Harko:2014pqa, Moraes:2016mlp}. It has been observed that non-minimally coupled curvature-matter scenarios can generate a substantial amount of comoving entropy during the universe's late stages of evolution. This points toward a potential interpretation of the energy exchange between curvature and matter sectors as being due to gravitationally induced particle creation within the framework of the  FLRW universe. Our analysis of observational data consistently supports the plausibility of such an interpretation for certain values of the model parameters ($\alpha, \beta, n, k$).\\
 
The rest of the paper is structured as follows. In Sec.\ \ref{Sec:II}, we explore the theoretical framework of non-minimally coupled $f(R)$ models within a flat FLRW universe. We derive the generic modified evolution equations describing the universe in the presence of non-minimal curvature-matter coupling. Two distinct coupled models are considered, and the corresponding equations for each model are derived. In Sec.\ \ref{Sec:III}, we discuss the methodology for analyzing the Pantheon + OHD + BAO datasets. This analysis aims to elucidate the temporal behavior of various cosmological quantities during the late-time phase of cosmic evolution. In Sec.\ \ref{Sec:IV}, we present the constraints on model parameters $k$ and $n$ under specific choices of coupling parameters. These constraints are derived from the analysis of observed data, and the results are thoroughly discussed. Finally, we summarize the conclusions of the paper in Sec.\ \ref{Sec:V}.

\section{Theoretical framework of curvature-matter coupling scenario}
\label{Sec:II}

In the context of modified gravity theories featuring non-minimal curvature-matter coupling, the action can be expressed as \cite{Harko:2008qz},

\begin{eqnarray}
 S &=& \int d^{4}x \sqrt{-g} \left[\frac{1}{2\kappa^2}f_1(R)+\mathcal{G}({\cal L}_{m})f_2(R)\right]\,,
\label{eq:b1}
\end{eqnarray}
Here, $f_1(R)$ and $f_2(R)$ are two  arbitrary functions of Ricci scalar $R$ and ${\cal L}_{m}$ denotes the matter Lagrangian density. $g$ is the determinant of the spacetime metric   $g_{\mu\nu}$ and $\mathcal{G}({\cal L}_{m})$ represents  matter Lagrangian-dependent coupling term, responsible for
 curvature-matter interactions. $G$ denotes the gravitational constant
 and   in this work, we have set   $\kappa^2 = 8\pi G = 1$.
 The action described above can be simplified into a minimally coupled curvature-fluid 
 scenario by setting $f_1(R)=R$, $f_2(R)=1$, and 
$\mathcal{G}({\cal L}_{m})={\cal L}_{m}$. On the other hand, one type of non-minimally coupled curvature-fluid scenario can be obtained from this framework by choosing $f_1(R)=R$, $f_2(R)=R^n$, and $\mathcal{G}({\cal L}_{m})=(1 + \lambda {\cal L}_{m})$ \cite{fr6}. In this work, we do not confine ourselves to specifying the functional form of $\mathcal{G}({\cal L}_{m})$ in advance, to obtain the field equations corresponding to the above action. Following
the approach of  metric formalism  within the context of modified gravity theories, the variation of the action with respect to    $g_{\mu \nu }$ yields the modified field equations as, 
\begin{equation}
\begin{split}
F_1(R)R_{\mu \nu } - \frac{1}{2}f_1(R)g_{\mu \nu} + (g_{\mu\nu}\square - \nabla_\mu \nabla_\nu) F_1(R) &= -2 \mathcal{G}({\cal L}_{m}) F_2(R) R_{\mu\nu} \\
& \quad - 2(g_{\mu\nu}\square - \nabla_\mu \nabla_\nu) F_2(R) \mathcal{G}({\cal L}_{m}) \\
& \quad - f_2(R)\left(\frac{\partial \mathcal{G}}{\partial \mathcal{L}_m}\mathcal{L}_m - \mathcal{G}({\cal L}_{m})\right)g_{\mu\nu} \\
& \quad + T_{\mu\nu} f_2(R) \frac{\partial \mathcal{G}}{\partial \mathcal{L}_m} \, ,
\end{split}
\label{eq:b2}
\end{equation}
where $F_i = df_i/dR$ with $i=1,2$ and the matter energy-momentum tensor $T_{\mu \nu}$ is expressed as,
 
\begin{eqnarray}
T_{\mu \nu}=-{2 \over \sqrt{-g}}{\delta(\sqrt{-g}{\cal
L}_m)\over \delta(g^{\mu\nu})} \,.
\label{eq:b3}
\end{eqnarray}

To investigate the   impact of the term $\Big{(}\frac{\partial \mathcal{G}}{\partial \mathcal{L}_m}\mathcal{L}_m - \mathcal{G}({\cal L}_m)\Big{)}$,
when it is non-vanishing, on modified field eq.\ (\ref{eq:b2}), we have considered two types of non-minimal coupling scenarios, namely (I) linear coupling, where  $\mathcal{G}({\cal L}_m) = \alpha {\cal L}_m - \beta$, and (II) non-linear coupling, where $\mathcal{G}({\cal L}_m) = \alpha{\cal L}_m \Big{(} \ln \Big{(}\frac{{\cal L}_m}{\beta} \Big{)}+ 1 \Big{)}$.  
The expression $\Big{(}\frac{\partial \mathcal{G}}{\partial \mathcal{L}_m}\mathcal{L}_m - \mathcal{G}({\cal L}_m)\Big{)}$
becomes a constant ($\beta$) for the linear case (I) and is equal to $\alpha {\cal L}_m$
in the non-linear case (II).\\

In this study, we explore the implications of both linear and non-linear non-minimal coupling between curvature and matter within the framework of pure Einstein gravity, where $f_1(R)=R$ and the matter component is non-minimally coupled (with coupling strength $\lambda$) to curvature through the function $f_2(R) = R^n$ (where $n$ is a constant). The matter content of the universe is modeled as a perfect fluid characterized by energy density $\rho$ and pressure $p$. Following   discussions elaborated in \cite{Schutz:1970, Brown:1993}, we adopt $\mathcal{L}_m = p$ as a natural choice for the Lagrangian density of perfect fluids, which accurately reproduces the hydrodynamical equations for such systems. This specific choice has significant implications in the examination of curvature-matter coupling scenarios, implying the absence of an additional force due to deviations from geodesic motion, which may arise from the non-vanishing covariant derivative of $T_{\mu\nu}$ within the context of curvature-matter coupling \cite{Bertolami:2007}. However, alternative choices for $\mathcal{L}_m$ have been explored in previous studies \cite{Bertolami:2008ab, Brown:1993}. To summarize our considerations for the present analysis, we adopt $f_1(R) =R$, $f_2(R) = R^n$, $F_2(R) = nR^{n-1}$, and ${\cal L}_m = p$.  
 To proceed with our investigation, we opted for an exponential functional form 
 $p = p_0e^{ak}$ for the pressure of the fluid sector, as detailed later in the study.
Evidently, this choice provides straightforward simplifications for the resulting equations involving the time derivatives of $p$ ($\dot{p},\ddot{p}$). 
The  covariant divergence of energy-momentum tensor (\ref{eq:b3}) gives
\begin{eqnarray}
\nabla^\mu T_{\mu\nu} &=& \nabla^\mu \ln \left[f_2(R) \frac{\partial \mathcal{G}(\mathcal{L}_m)}{\partial \mathcal{L}_m}\right] (\mathcal{L}_m g_{\mu\nu} - T_{\mu\nu}) = Q_{\nu}
\label{eq:b4}
\end{eqnarray}
The choice $\mathcal{L}_m = p$ ensures presence of non-zero source term $Q_{\nu}$
in eq.\ (\ref{eq:b4}) for both types of couplings mentioned above.\\

\subsection*{Case I: Linear coupling} The linear type coupling described by the choice $\mathcal{G}({\cal L}_{m}) = \alpha {\cal L}_{m} - \beta$  results in 
 modification of the field eq.\ (\ref{eq:b2}) as follows,
\begin{eqnarray}
R_{\mu \nu }- {1\over 2}Rg_{\mu \nu }
&=&  -2 (\alpha \mathcal{L}_m - \beta)(nR^{n-1})R_{\mu\nu} \nonumber\\ 
&& -2(g_{\mu\nu}\square - \nabla_\mu \nabla_\nu) (nR^{n-1}) (\alpha \mathcal{L}_m - \beta) + R^n(\alpha T_{\mu\nu} - \beta g_{\mu\nu}) 
\label{eq:b5}
\end{eqnarray}
The  expressions for energy density $\rho$ and pressure $p$ of the fluid 
can be obtained from the $00$  and $ii$  components of the energy-momentum tensor $T_{\mu\nu}$. The  $00$  component of eq.\ (\ref{eq:b4}) gives as, 
\begin{eqnarray}
R_{00} - {1\over 2}Rg_{00}
&=&  -2 (\alpha \mathcal{L}_m - \beta)(nR^{n-1})R_{00} \nonumber\\ 
&& -2(g_{00}\square - \nabla_\mu \nabla_\nu) (nR^{n-1}) (\alpha \mathcal{L}_m - \beta) + R^n(\alpha T_{00} - \beta g_{00}) 
\label{eq:b6}
\end{eqnarray}
For a FLRW spacetime background, the above equation can be reformulated to yield
\begin{eqnarray}
\rho &=& \frac{1}{\alpha R^n}\left[3H^2 + 6(\beta - \alpha p)(nR^{n-1})(H^2 + \dot{H}) + 6H(n(n-1)R^{n-2}\dot{R}(\alpha p - \beta)+nR^{n-1}\alpha \dot{p})-\beta R^n\right]\nonumber\\
\label{eq:b7}
\end{eqnarray}
On the other hand, the `$ii$'-component of eq.\ (\ref{eq:b4}) gives,
\begin{eqnarray}
R_{11}- {1\over 2}Rg_{11}
&=& 
-2 (\alpha \mathcal{L}_m - \beta)(nR^{n-1})R_{11} \nonumber\\ 
&& -2(g_{11}\square - \nabla_\mu \nabla_\nu) (nR^{n-1}) (\alpha \mathcal{L}_m - \beta) + R^n(\alpha T_{11} - \beta g_{11})\,,
\label{eq:b8} 
\end{eqnarray}
which in a FLRW background can be expressed as  
\begin{eqnarray}
- (2\dot{H} + 3H^2) &=& p\Big{[}-2\alpha(nR^{n-1})(\dot{H} + 3H^2)+\alpha R^n + 4Hn(n-1)R^{n-2}\dot{R}\alpha \nonumber\\
&& + 2n(n-1)(n-2)R^{n-3}\dot{R}^2\alpha + 2n(n-1)R^{n-2}\ddot{R}\alpha\Big{]}\nonumber\\
&& + \dot{p}\left(4HnR^{n-1}\alpha + 4n(n-1)R^{n-2}\alpha \dot{R}\right)  + \ddot{p}(\alpha R^{n-1}n) \nonumber\\
&& + \beta \Big{[}2nR^{n-1}(\dot{H} + 3H^2) - R^n - 4Hn(n-1)R^{n-2}\dot{R}\nonumber\\
&& - 2n(n-1)(n-2)R^{n-3}\dot{R}^2 - 2n(n-1)R^{n-2}\ddot{R}\Big{]} 
\label{eq:b9} 
\end{eqnarray}
Eqs.\ (\ref{eq:b7}) and (\ref{eq:b9})  serve as the master equations for our analysis within the framework of curvature-matter coupling in a FLRW spacetime background. Notably, these equations incorporate time derivatives of the fluid pressure, reflecting the influence of non-minimal curvature-matter coupling on cosmic evolution.
When both the non-minimal coupling constants $\alpha, \beta$ are set to zero in these   equations, the standard Friedmann equations are recovered, 
which do not involve any derivative terms of pressure. \\

To explore the observational constraints on curvature-matter coupling models, we examine fluid pressure models characterized by specific temporal profiles. 
To describe the temporal profiles of fluid pressure 
$p$, we adopt an exponential form expressed in terms of the FLRW scale factor  $a$. Specifically,  $p$ is given by  $p=p_0\exp(ak)$, where $k$ is a dimensionless parameter and $p_0$ is a constant  representing pressure of the fluid at the present epoch ($a=1$).
Consequently, for this model,
 we have $\dot{p} = kp\dot{a}$ and $\ddot{p} = k^2p\dot{a}^2 + kp\ddot{a}$ 
and  we can express the energy density  in eq.\ (\ref{eq:b7}) 
for a given set of parameters $\alpha,\beta,n,k$, and pressure $p$ as,  
\begin{eqnarray}
\rho(t;\alpha,\beta,n,k) &=& \frac{1}{\alpha R^n}\Big{[}3H^2 +6(\beta - \alpha p)(nR^{n-1})(H^2 + \dot{H})\nonumber\\
&& + 6H\left(n(n-1)R^{n-2}\dot{R}(\alpha p - \beta) + nR^{n-1}\alpha k p H a\right)-\beta R^n\Big{]}
\label{eq:b10}
\end{eqnarray}
Similarly using eq.\ (\ref{eq:b9}) we can express the pressure $p$  as 
\begin{eqnarray}
p(t;\alpha,\beta,n,k) &=& -\Big{[}(2\dot{H} + 3H^2) + \beta \big{(}2nR^{n-1}(\dot{H} + 3H^2) - R^n -4Hn(n-1)\ddot{R}R^{n-2}\nonumber \\
&& -2n(n-1)(n-2)R^{n-3}\dot{R}^2 - 2n(n-1)R^{n-2}\ddot{R}\big{)}\Big{]}\Biggl{/}\Big{[}-2\alpha (nR^{n-1})(\dot{H} + 3H^2)\nonumber \\
&& + \alpha R^n + 4Hn(n-1)R^{n-2}\alpha \dot{R} + 2n\alpha(n-1)(n-2)R^{n-3}\dot{R}^2 + 2n\alpha (n-1)R^{n-2} \ddot{R}\nonumber \\
&& + \dot{a}k\left(4HnR^{n-1}\alpha + 4n(n-1)\alpha R^{n-2}\dot{R}\right) + (k\dot{a}^2+k\ddot{a})(\alpha n R^{n-1})\Big{]}
\label{eq:b11}
\end{eqnarray}
For the selected form of linear coupling, the divergence of the energy-momentum tensor yields a non-zero source term. The `$0$' component of the source term in eq.\  (\ref{eq:b4})  
can be represented as,
\begin{eqnarray}
Q_{0} = -\frac{\alpha n \dot{R}}{R} (\rho + p)
\label{eq:b12}
\end{eqnarray}
The  source term $Q_0$ depends on the specific values of the model parameters $\alpha, n, k$, and also on the specific pressure model chosen. It plays a crucial role in determining the evolutionary dynamics within the framework of the selected curvature-matter coupling model.

\subsection*{Case II: Non-Linear   coupling } 
In this coupling scenario described by  
$\mathcal{G}({\cal L}_{m}) = \alpha  {\cal L}_{m} \Big{(} \ln \Big{(}\frac{{\cal L}_{m}}{\beta} \Big{)}+ 1 \Big{)}$,
  the coupling takes a non-linear (logarithmic) 
  form  of the matter Lagrangian (${\cal L}_{m}$) 
  with coupling coefficients $\alpha$ and $\beta$. This kind of coupling   leads to the following modification of the field eq.\ (\ref{eq:b2}),
\begin{eqnarray}
R_{\mu \nu }- {\frac12}Rg_{\mu \nu }
&=&  -2\alpha ( p \ln (p/\beta) +  p)(nR^{n-1})R_{\mu\nu} \nonumber\\ 
&& -2\alpha(g_{\mu\nu}\square - \nabla_\mu \nabla_\nu) (nR^{n-1}) ( p \ln (p/\beta) +   p)\nonumber\\ 
&& + R^n\left[(\alpha  \ln (p/\beta) + 2 \alpha) T_{\mu\nu} - \alpha p g_{\mu\nu}\right]
\label{eq:c1}
\end{eqnarray}
%
%The `$00$' component of eq.\ (\ref{eq:c1}) gives
%%
%\begin{eqnarray}
%R_{00} - {\frac12} Rg_{00}
%&=&  -2 (\alpha p \ln (p/\beta) +  \alpha p)(nR^{n-1})R_{00} \nonumber\\ 
%&& -2(g_{00}\square - \nabla_\mu \nabla_\nu) (nR^{n-1}) (\alpha p \ln (p/\beta) +  \alpha p)\nonumber\\ 
%&& + R^n\left[(\alpha  \ln (p/\beta) + 2 \alpha) T_{00} - \alpha p g_{00}\right] 
%\label{eq:c2}
%\end{eqnarray}
%%
The `$00$' component of $T_{\mu\nu}$ gives the 
energy density  $\rho$  and using the `$00$' component of
the above equation  for a FLRW background,
$\rho$ can be expressed as
\begin{eqnarray}
\rho &=& \frac{1}{(\alpha  \ln (p/\beta) + 2 \alpha) R^n}\Big{[}3H^2 - 6(\alpha p \ln (p/\beta) +  \alpha p)(nR^{n-1})(H^2 + \dot{H})\nonumber\\ 
&& + 6H\left(n(n-1)R^{n-2}\dot{R}(\alpha p \ln (p/\beta) +  \alpha p)+nR^{n-1}(\alpha \dot{p} \ln (p/\beta) + 2 \alpha \dot{p})\right)\nonumber\\ 
&& -\alpha p R^n\Big{]}
\label{eq:c3}
\end{eqnarray}
On the other hand, the `$ii$' component of $T_{\mu\nu}$ gives the pressure $p$
and for  the `$00$' component of the eq.\ (\ref{eq:c5}) for a FLRW spacetime 
background gives the equation
\begin{eqnarray}
- (2\dot{H} + 3H^2) &=& p\Big{[}-2(\alpha \ln (p/\beta) + \alpha)(nR^{n-1})(\dot{H} + 3H^2)+(\alpha \ln (p/\beta) + \alpha) R^n \nonumber\\
&& + 4Hn(n-1)R^{n-2}\dot{R}(\alpha \ln (p/\beta) + \alpha)  \nonumber\\
&& + 2n(n-1) (\alpha \ln (p/\beta) + \alpha) \left((n-2)R^{n-3}\dot{R}^2 + R^{n-2}\ddot{R} \right)\Big{]}
\nonumber\\
&& + \dot{p}\left(4HnR^{n-1}(\alpha \ln (p/\beta) + 2\alpha) + 4n(n-1)R^{n-2}(\alpha \ln (p/\beta) +2 \alpha) \dot{R}\right) \nonumber\\
&&  + 2\ddot{p}((\alpha \ln (p/\beta) +  \alpha) nR^{n-1}) + 2n R^{n-1} \alpha  \frac{\dot{p}^2}{p} 
\label{eq:c5} 
\end{eqnarray}
Similar to the previous case, we  consider the fluid pressure model with $p=p_0\exp(ak)$,   and for simplicity, we set $\beta=p_0$. 
At present epoch ($a=1$) the pressure of the fluid is given by
$p_0e^k$  Consequently, using $\dot{p} = kp\dot{a}$ and $\ddot{p} = k^2p\dot{a}^2 + kp\ddot{a}$ in eqs. (\ref{eq:c3}) and (\ref{eq:c5}), we can express the energy density $\rho$
and pressure $p$ for a given parameter set  $\alpha,\beta,n,k$ respectively as 
\begin{eqnarray}
\rho(t;\alpha,\beta,n,k) &=& \frac{1}{(\alpha a k+ 2 \alpha) R^n}\Big{[}3H^2 -6\alpha p( a k + 1)(nR^{n-1})(H^2 + \dot{H})\nonumber\\
&& + 6H\left(\alpha pn(n-1)R^{n-2}\dot{R}( a k+ 1) + nR^{n-1}( a k+  1) \alpha k p Ha\right)
 -\alpha p R^n\Big{]}\nonumber\\
\label{eq:c6}
\end{eqnarray}
and
\begin{eqnarray}
p(t;\alpha,\beta,n,k) &=& - \left(2\dot{H} + 3H^2\right)\Biggl{/}\Big{[}-2\alpha( a k + 1)nR^{n-1}(\dot{H} + 3H^2) + \alpha R^n ( a k + 1) \nonumber\\
&& + 4H  \left(\alpha( a k + 1)(n(n-1)R^{n-2}\dot{R})+(nR^{n-1})( a k + 2)\alpha k\dot{a}\right)\nonumber\\
&& + 2 \Big{[}\left(n(n-1)(n-2)R^{n-3}\dot{R}^2 + n(n-1)R^{n-2}\ddot{R}\right)( a k + 1)\alpha \nonumber\\
&& + 2(n(n-1)R^{n-2}\dot{R})( a k + 2)\alpha k\dot{a} \nonumber\\
&& + nR^{n-1} \alpha \left(( a k  + 2)(k^2\dot{a}^2 + k\ddot{a}) + \alpha k^2 H^2 a^2\right)\Big{]}
\label{eq:c7}
\end{eqnarray} 
For this non-linear coupling scheme, the divergence of the energy-momentum 
tensor involves a non-zero source term whose `$0$' component is given by
\begin{eqnarray}
Q_{0} 
&=&
-\frac{\alpha \big{(} nR^{n-1} \dot{R}(ak+2) + R^n k\dot{a}\big{)}}{R^n(ak+2)} (\rho + p)
\label{eq:c8}
\end{eqnarray}
 Besides the model parameters $(n, k)$  and coupling parameters $(\alpha, \beta)$,
the expressions for $\rho$ and $p$ in both types of coupled models also depend on additional cosmological parameters. These include the FLRW scale factor $a$
and its time derivatives, the Hubble parameter $H$ and its time derivatives, 
the Ricci scalar$R$ ( = $6\dot{H}+12H^2$) and its time derivative as well.\\

We obtained the temporal profile of all these cosmological quantities during the late-time phase of cosmic evolution by analyzing combined observational data sets (Pantheon + OHD + BAO). This analysis covers a redshift  ($z$) range from $0$ to $2.3$, corresponding to a cosmic time domain of $0.23$ to $1$, with $t$ normalized to unity at the present epoch.  
Using these profiles in eqs.\ (\ref{eq:b10}), (\ref{eq:b11}), (\ref{eq:c6}), (\ref{eq:c7}),
we determined the time evolution of the energy density and pressure of the fluid under both linear and nonlinear coupling scenarios.  We constrained the model parameters ($k, n$) using various classical energy conditions. 
We have also investigated the time evolution of  effective equation of state (EoS)
parameter for both the coupling scenarios for selected benchmark points 
selected from the allowed domain of the ($k, n$) parameter space.

\section{Constraints on cosmological quantities from the combined analysis of observational data}
\label{Sec:III}

Cosmological data  compiled from Supernovae Ia observations offer valuable insight into the cosmic evolution during later epochs. Here, we provide a concise overview of the technical aspects of our analysis of the observational data, which comprises the Pantheon dataset (SNe Ia) with 1048 data points \cite{Pan-STARRS1:2017jku}, Observed Hubble Data (OHD) with 54 data points \cite{Geng:2018pxk}, and Baryon Acoustic Oscillation (BAO) data \cite{WiggleZ}. By employing the Markov Chain Monte Carlo (MCMC) technique with the emcee package as the optimizer \cite{emcee}, implemented within the lmfit library in Python \cite{lmfit}, we explain how we extracted the temporal behavior of the FLRW scale factor $a$ and its derivatives, which are crucial for constraining non-minimally coupled matter-curvature models.\\

We utilized uniform priors for the parameters, with Hubble parameter at present epoch \(H_0\) ranging from $65 - 75$ \(\text{km s}^{-1} \text{Mpc}^{-1}\) and dark-matter density at present epoch $(\Omega_{m0})$ ranging from 0.1 to 0.5. The best-fit values were determined by minimizing the \(\chi^2\) function as described in \cite{John:2023fsy},
\begin{eqnarray}
\chi^2 &=& \sum_{i=1}^N\Big[\frac{\xi_i^{obs} - \xi_i^{th}}{\sigma_i}\Big]^2
\label{eq:O1}
\end{eqnarray}
Here, $\xi_i^{obs}$ represents the observed value from the datasets, $\xi_i^{th}$ stands for the corresponding theoretical value derived from our model, and $\sigma_i$ denotes the error associated with the measured values. We define the reduced $\chi^2$, denoted as $\chi^2_{\rm d.o.f}$, as the minimum $\chi^2$ per degrees of freedom, calculated as $\chi^2_{\rm d.o.f} = \frac{\chi^2}{N-N_p}$, where $N$ represents the total number of data points and $N_p$ denotes the number of  parameters. Additionally, we incorporate the Akaike Information Criterion (AIC) and Bayesian Information Criterion (BIC) into our analysis, defined as follows,
\begin{eqnarray}
\rm AIC &=& N \ln (\chi^2/N) + 2N_p \label{eq:o2} \\
\rm BIC &=& N \ln (\chi^2/N) + \ln(N)N_p \label{eq:o3}
\end{eqnarray}

Similar to $\chi^2$, models with lower AIC and BIC values are preferred. Below, we list the observations considered for the estimation.

\begin{itemize}
\item \textbf {OHD data-sets:}
This dataset contains 52 data points of $H(z)$ measurements covering a redshift range from $0 
\leq z \leq 2.36$. It includes 31 data points obtained using the differential age (DA) method 
and 20 from clustering measurements \cite{Geng:2018pxk}.  
The DA method, introduced by Jimenez and Loeb \cite{Jimenez:2001gg}, 
determines the Hubble parameter by comparing the ages of passively-evolving galaxies that have similar metallicity and are spaced closely in redshift. 
Clustering measurements, initially proposed by Gaztañaga et al. \cite{Gaztanaga:2008xz}, use the BAO peak positions as a standard ruler along the radial direction by analyzing the clustering patterns of galaxies or quasars. Additionally, one data point comes from a direct measurement of the Hubble constant, \(H_0 = 73.24 \pm 1.74 \, 
\text{km/s/Mpc}\), based on observations from the Hubble Space Telescope 
\cite{Riess:2016jrr}.

\item \textbf {SNe Ia data-sets}: 
This dataset utilizes the most recent Pantheon sample \cite{Pan-STARRS1:2017jku}, which comprises luminosity data from 1048 Type Ia supernovae across a redshift range of $0.01<z<2.3$. Scolnic et al. \cite{Pan-STARRS1:2017jku} employed the SALT2 light curve fitter \cite{SN} to determine the observed distance modulus.
\begin{eqnarray}
\mu_i^{obs} = m(z)  + \alpha  X_1 - \beta  C - M
\label{eq:o4}
\end{eqnarray}
The luminosity distance for a Type Ia supernova event at redshift $z_i$ can be calculated using the following equation,
\begin{eqnarray}
d_L(H_0,\Omega_{m}^0,z_i) = c(1+z_i) \int_{0}^{z_i}\frac{dz'}{H(H_0,\Omega_{m}^0,z')} \,,
\label{eq:o5}
\end{eqnarray}
where,  $c$ represents the speed of light, and $H(H_0,\Omega_{m}^0,z')$ refers to the Hubble parameter as defined by the model. The theoretical distance modulus is then given by the following expression,
\begin{eqnarray}
\mu_i^{th}(H_0,\Omega_{m}^0,z_i) = 5\log_{10}\Big[\frac{d_L(H_0,\Omega_{m}^0,z_i)}{Mpc}\Big]+25.
\label{eq:o6}
\end{eqnarray}

\item \textbf {BAO data-sets:}
The BAO data is employed to determine the acoustic parameter $A$, which describes the distance-redshift relationship from a baryon acoustic oscillation (BAO) measurement. The theoretical acoustic parameter is formulated using the model parameters as follows \cite{John:2023fsy},
\begin{eqnarray}
A(H_0,\Omega_{m}^0) = \frac{\sqrt{\Omega_{m}^0}}{h(z_A)^{1/3}}\Big(\frac{1}{z_A}\int_{0}^{z_A}\frac{dz}{h(H_0,\Omega_{m}^0,z)}\Big)^{2/3}
\label{eq:o7}
\end{eqnarray}
To compute $\chi_{\rm BAO}^2$, we compare this theoretical value with the observed value $A_{\rm obs}=0.484\pm0.016$ at redshift $z_A=0.35$, derived from SDSS-BAO distance measurements \cite{WiggleZ}.	
\end{itemize}

\begin{figure}[H]
\centering
\includegraphics[width=0.7\textwidth, height=4.5in]{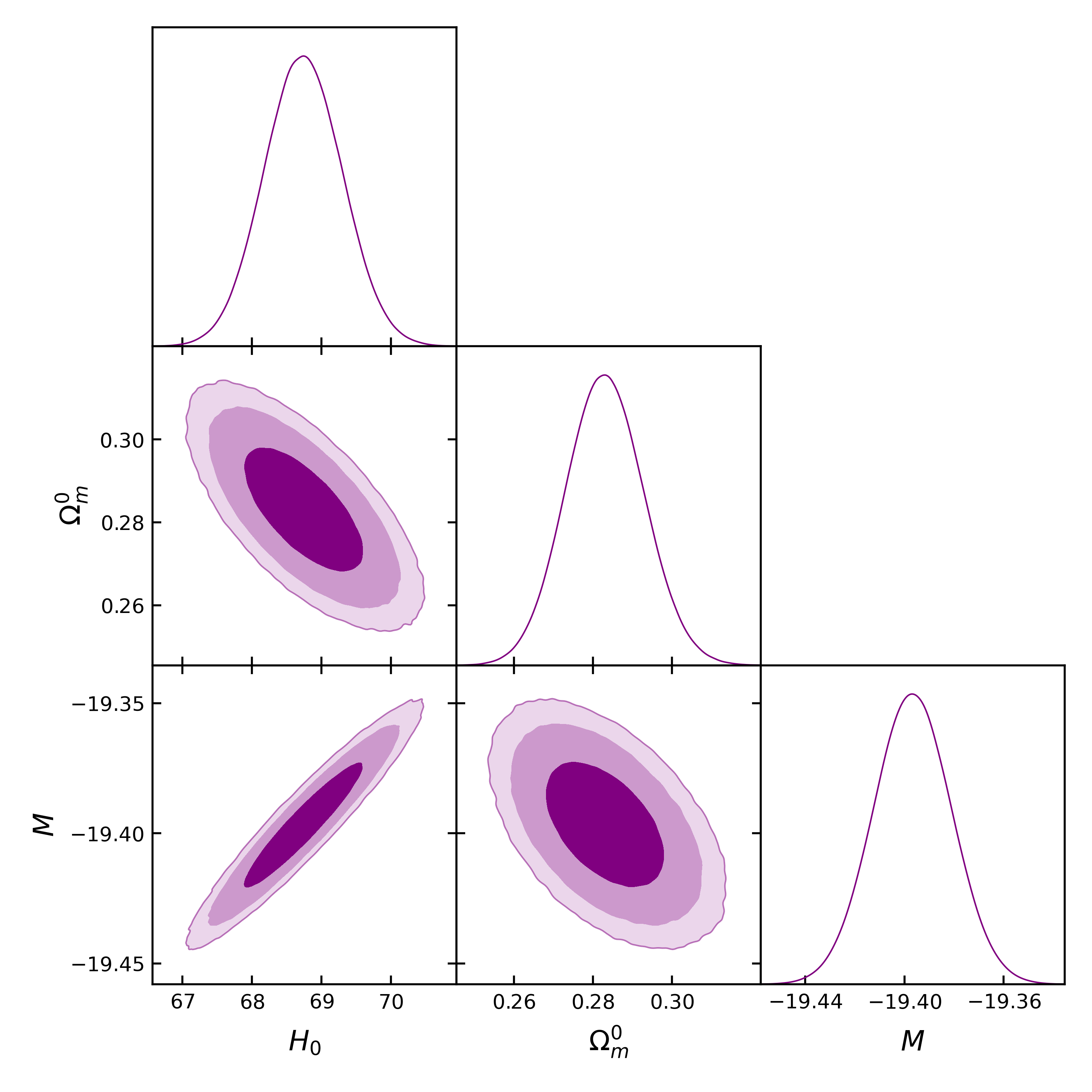}
\caption{Posterior distribution plot of $H_0, \Omega^0_{m}$ and $M$}
\label{fig:F1}
\end{figure}

To estimate the parameters \( H_0 \), \( \Omega^0_{m} \), and \( M \)  along with their uncertainties, we employ Markov chain Monte Carlo (MCMC) Bayesian parameter estimation using uniform priors for all parameters. We utilize the publicly available Python packages emcee \cite{Goodman} for generating MCMC samples and GetDist \cite{Lewis} for plotting the posterior distributions of
\( H_0 \), \( \Omega^0_{m} \), and \( M \).
After assessing the independence and convergence of the MCMC samples and thinning them as necessary, we derive the following estimates from a combined analysis of Pantheon data, OHD, and BAO. Fig.\ (\ref{fig:F1}) displays the posterior distribution plots for highlighting the corresponding 1-$\sigma$ and 3-$\sigma$ uncertainties for these parameters.

\begin{eqnarray}
\Omega^0_{m} = 0.28^{+0.009}_{-0.009},\quad M = -19.39^{+0.015}_{-0.015} , \quad {\rm and} \quad  H_0 = 68.74^{+0.56}_{-0.56} \mbox{ km s$^{-1}$ Mpc$^{-1}$}\,.
\label{eq:o8} 
\end{eqnarray}

The combined $\chi^2$ value for the datasets
is calculated by adding the individual minimum $\chi^2$
values from each dataset, resulting in a total 
$\chi^2_{\rm d.o.f}$ value of \(0.973\).
Furthermore, the Akaike Information Criterion (AIC) and the Bayesian Information Criterion (BIC) are computed to assess the model's suitability for the combined datasets. The AIC for the combined data is \(-26.708\), and the BIC is  \(-11.71\). These criteria provide a means to assess the relative quality of the statistical models by balancing goodness of fit and model complexity.  These metrics serve to evaluate the relative effectiveness of the statistical models, weighing the trade-off between goodness of fit and model complexity. Models with lower AIC and BIC values are considered more effective, as they provide a better fit to the data while efficiently using fewer parameters.

\begin{figure}[H]
\centering
\begin{subfigure}[b]{0.51\textwidth}
\includegraphics[width=\textwidth]{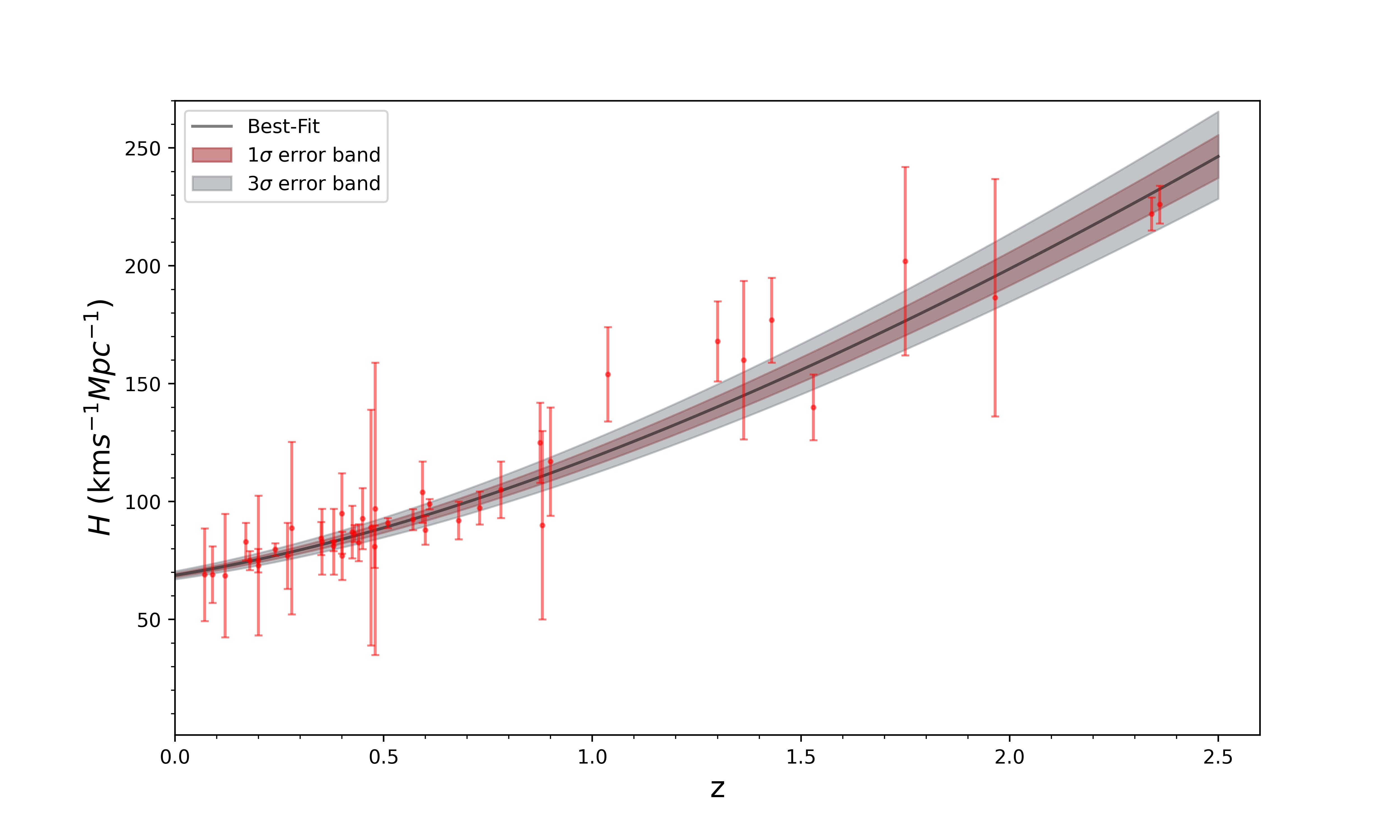}
\end{subfigure}
\begin{subfigure}[b]{0.46\textwidth}
\includegraphics[width=\textwidth]{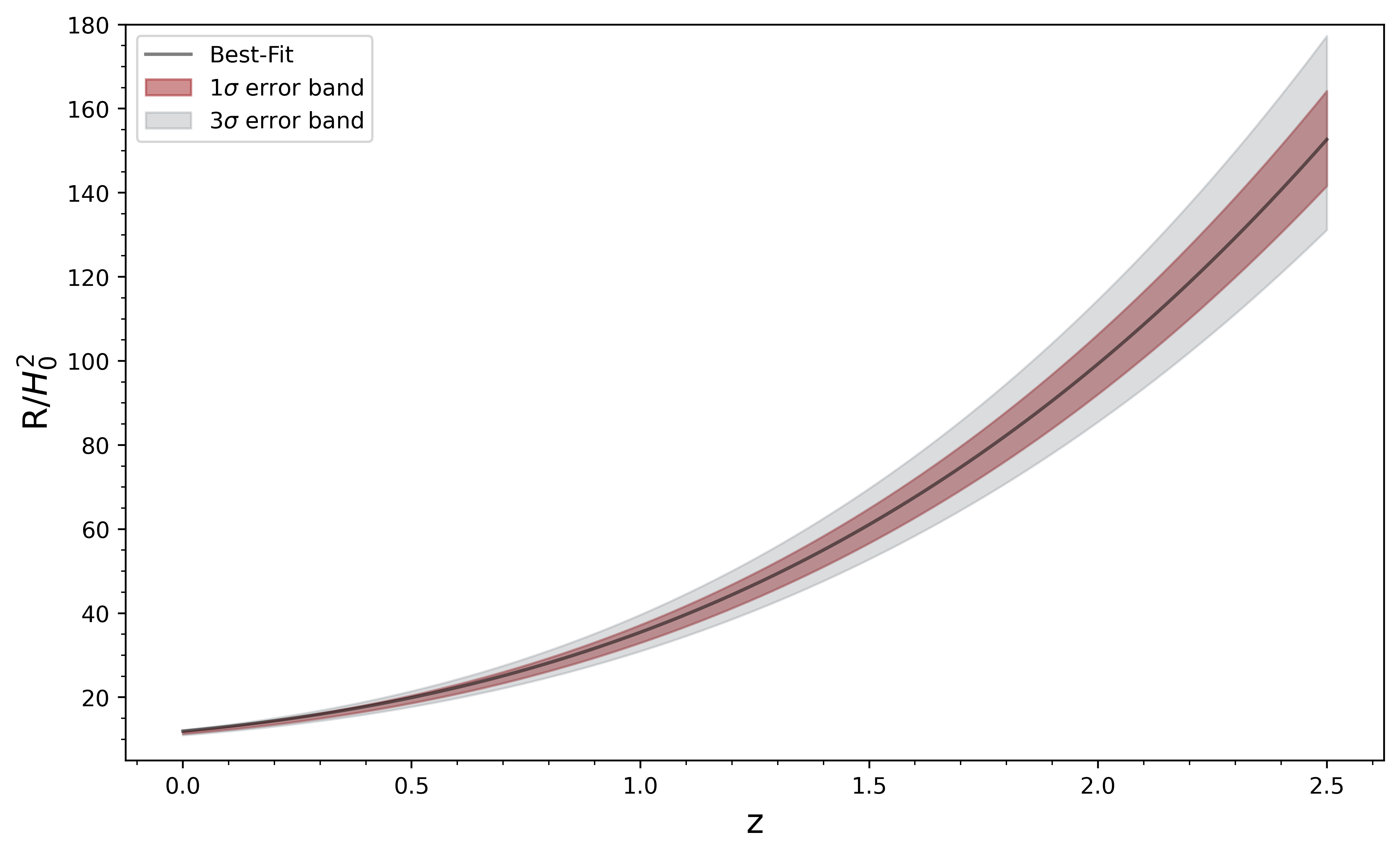}
\end{subfigure}
\caption{Left Panel: Best-fit curve with 1-$\sigma$ and 3-$\sigma$ uncertainties obtained for $H$ vs z.  Right Panel: Best-fit curve along with 1-$\sigma$ and 3-$\sigma$ uncertainties determined for $R/H_0^2$ vs z.} 
\label{fig:F2}
\end{figure}

The obtained best-fit values and uncertainties of the parameters \( H_0 \), \( \Omega^0_{m} \), and \( M \), are utilized to determine the redshift dependencies of the Hubble parameter \(H\) and the normalized Ricci Scalar \(R/H_0^2\), with  \(R = 12H^2 + 6\dot{H}\). The plots for \(H\) and \(R/H_0^2\) versus \(z\) are displayed in the left and right panels of    fig.\ (\ref{fig:F2})  respectively. The solid lines in fig.\ (\ref{fig:F2}) represent the best-fit curves in each case. The \(1\sigma\) and \(3\sigma\) uncertainties in the derived dependencies are indicated by dashed lines.

\begin{figure}[H]
\centering
\begin{subfigure}[b]{0.49\textwidth}
\includegraphics[width=\textwidth]{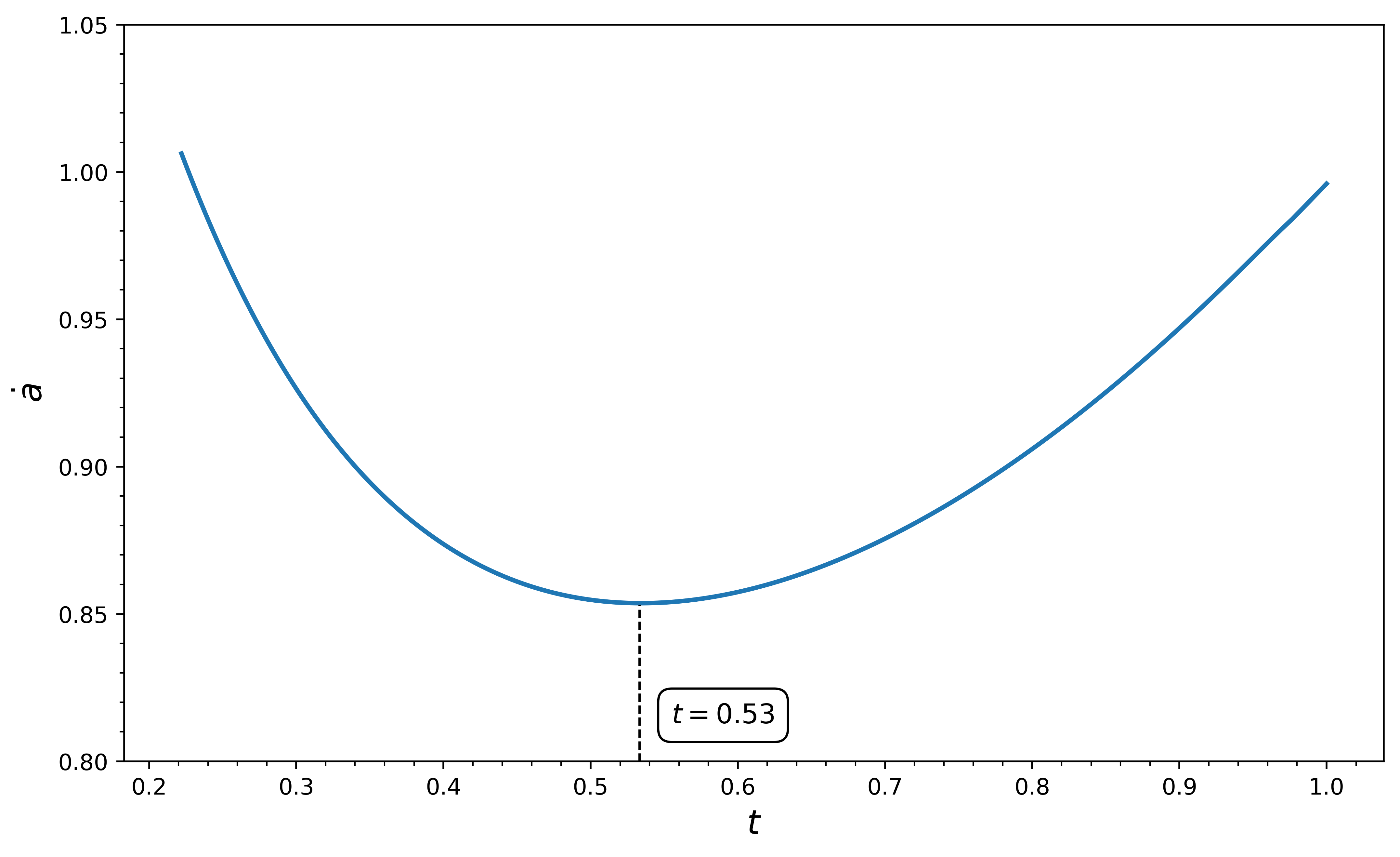}
\end{subfigure}
\begin{subfigure}[b]{0.49\textwidth}
\includegraphics[width=\textwidth]{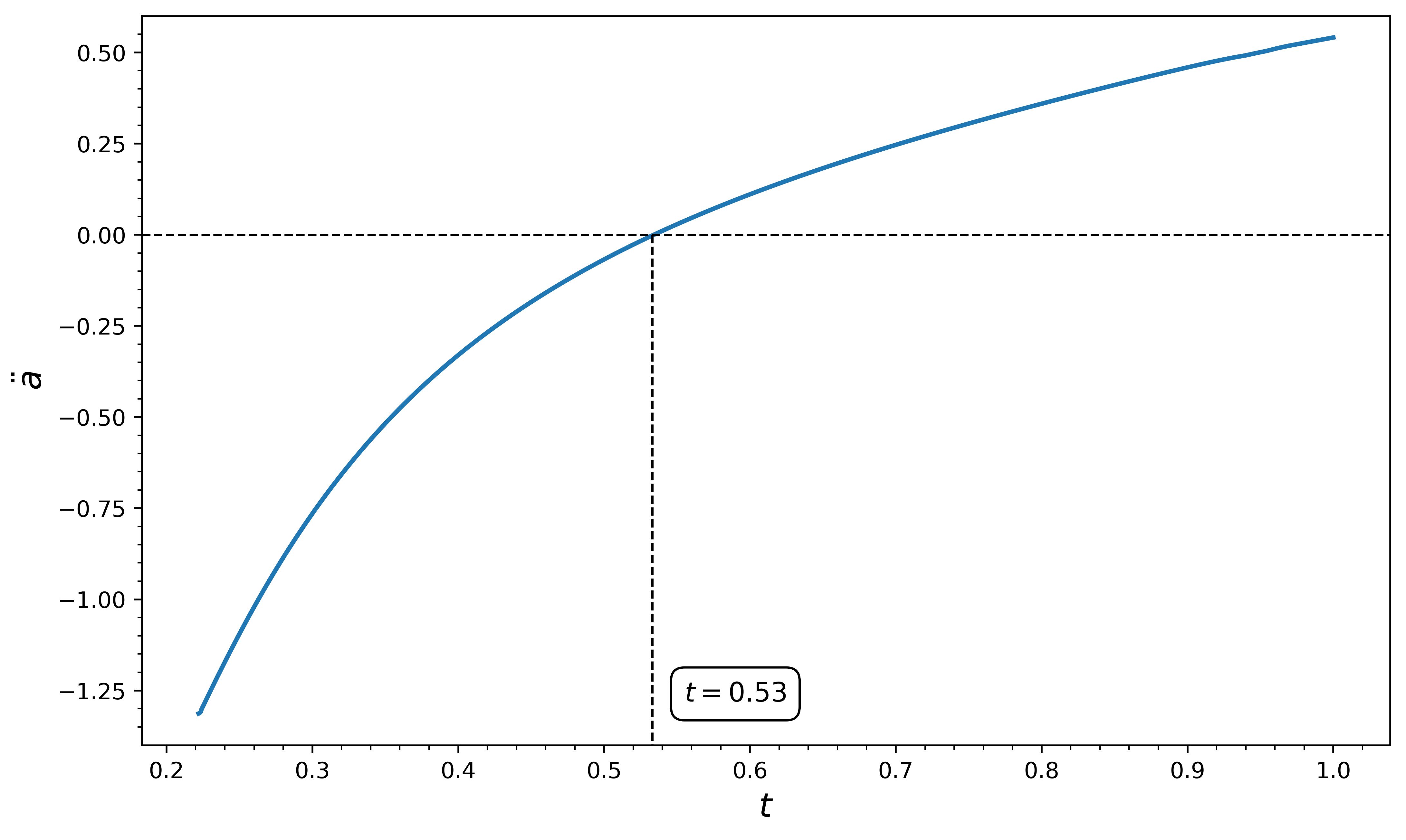}
\end{subfigure}
\caption{Variation of $\dot{a}$ and $\ddot{a}$ with cosmic time ($t$) at the best-fit parameter values obtained from analysis of combined data-sets.} 
\label{fig:F3}
\end{figure}

The profile of the \(H(z)\) function, derived from analyzing observational data, can be utilized to ascertain the temporal progression of the FLRW scale factor \(a(t)\) and its  derivatives  \(\dot{a}\) and \(\ddot{a}\). The methodology for this computation is outlined briefly below. The FLRW scale factor \(a\) is normalized to unity at the current epoch (\(z=0\)) and is inversely related to the redshift  $1/a = 1+z$. Given that, \(H = \dot{a}/a\) it follows that \(dt = -\frac{dz}{(1+z)H(z)}\) and integrating this relationship  yields,
\begin{eqnarray}
\frac{t(z)}{t_0} &=& 1 - \frac{1}{ t_0}\int_{z}^0 \frac{dz'}{(1+z')H(z')}
\label{eq:o9}
\end{eqnarray}
We calculate $ t(z)/t_0 $ using eq.\ (\ref{eq:o9}), where the current epoch is normalized to unity ($ t_0 = 1 $). This calculation involves numerically integrating the  $H(z)$  profile shown in the left panel of fig.\ (\ref{fig:F2}). Through these calculations, we derive simultaneous values of  $a$ and $t$ for any given $z$
by utilizing the relation $1/a = 1+z$.
By varying $z$ from 0 (current epoch)  to about 2.4 (as relevant for the Pantheon dataset)  
 we  compute $t(z)$ and $a(z) = 1/(1+z)$
 in increments of $\Delta z = 0.01$.
This approach maps out the temporal evolution
of $a(t)$ over $0.23 < t < 1$, with $a$ normalized to 1 at $z=0$ (or $t=1$). From these results, we determine $\dot{a}(t)$ and $\ddot{a}(t)$   through numerical differentiation. These are plotted in the left and right panels of fig.\ (\ref{fig:F3}). The minimum at $t\sim 0.53$  in the  $\dot{a}$ profile (left panel) and the sign change in $\ddot{a}$ (right panel)  mark the transition from decelerated to accelerated phase of expansion during the late time phase of evolution.

 \section{Results and Discussion}
\label{Sec:IV}
 
In this section, we examine the observational constraints on cosmological parameters derived from data analysis. Subsequently, we present the bounds on the parameters of non-minimally coupled curvature-matter models using these constraints. We explore two scenarios of non-minimal coupling: linear and non-linear types, as previously described in Sec. \ref{Sec:II}.
The energy density of the coupled sectors for the two scenarios is defined by 
eq.\ (\ref{eq:b7}) and eq.\ (\ref{eq:c3}), respectively.
 In these equations, the temporal behavior of the energy density, 
 $\rho(t; \alpha,\beta, n,k)$,
 is expressed as a function of cosmological quantities such as the scale factor $a$, its derivative $\dot{a}$, the Hubble parameter $H$, the Ricci scalar $R$, and  the model parameters: $\alpha,\beta, n,k$. The temporal behavior of cosmological quantities  
 ($a,\dot{a},H,R$) during the late time phase of cosmic evolution has been derived from observational data and is discussed in section \ref{Sec:III}.
 The parameters $\alpha, \beta$ determine the strength of the non-minimal coupling between the interacting sectors. 
 The parameter $k$ is involved in the selected model for fluid pressure,
 whereas  $n$  denotes the exponent in the gravitational term
 $f_2(R) = R^n$, which is coupled to the matter Lagrangian.  
We applied several classical energy density conditions—namely, the Strong Energy Condition (SEC), Dominant Energy Condition (DEC), and Weak Energy Condition (WEC) - to constrain the model parameters $k$ and $n$ for   specific values of the coupling parameters 
($\alpha,\beta$).
We performed these analyses over the entire accessible time domain for $t$,
evaluating the cosmological quantities $( a, \dot{a}, H, R )$
at each $t$, as derived from observational data analysis. 
We present the obtained observational constraints on the parameters  
($\alpha, \beta, n, k$) by illustrating the permissible regions within the $k-n$
parameter space for specified values of the coupling parameters $\alpha, \beta$.
This is done across both the coupling scenarios: linear and non-linear types,
considered in this work.\\

In figs.\ (\ref{fig:F4}) and (\ref{fig:F5})
we illustrate the allowed domains in the
$k-n$ parameter space, derived from observational data, for three different values of the coupling parameters $\alpha$ and $\beta$ in both linear and non-linear coupling scenarios, respectively.  
Based on the  criteria of energy density and pressure profiles,
three distinct regions are identified 
within these parameter spaces.  We refer to the three distinct regions
as $C_1$, $C_2$ and $C_3$ in the text. $C_1$ corresponds to   the region
in $n-k$ parameter space for which the energy density $\rho$
remains non-negative at all accessible times, thereby validating the Weak Energy condition (WEC) 
($\rho \geq 0$). $C_2$ corresponds to the region where both $\rho \geq 0$ and $P \leq 0$ are satisfied, indicating an effective equation of state parameter $\omega_{\text{eff}} = \frac{P}{\rho} \leq 0$. This region simultaneously fulfills the conditions for the Null Energy Condition (NEC), Weak Energy Condition (WEC), and Dominant Energy Condition (DEC) with $\rho + P \geq 0$. Effective EoS parameter in this region lies between $0$ and $-1$. Finally, $C_3$ corresponds to the region where the conditions $\rho \geq 0$, $\rho + 3P \leq 0$, and $\rho + P \geq 0$ are simultaneously satisfied, resulting in an effective equation of state parameter $\omega_{\text{eff}}$ lying between $-\frac{1}{3}$ and $-1$. Therefore, the $C_3$ region indicates a non-phantom type of dark energy-dominated era of the universe and satisfies the WEC and NEC but violates the Strong Energy Condition (SEC). Below, we provide a detailed discussion of the obtained constraints on the model parameter space for both linear and non-linear coupling scenarios.

\subsection{Linear-coupling case:}
\begin{figure}[H]
  \centering
  \begin{tabular}{cc}
    \includegraphics[width=0.43\textwidth]{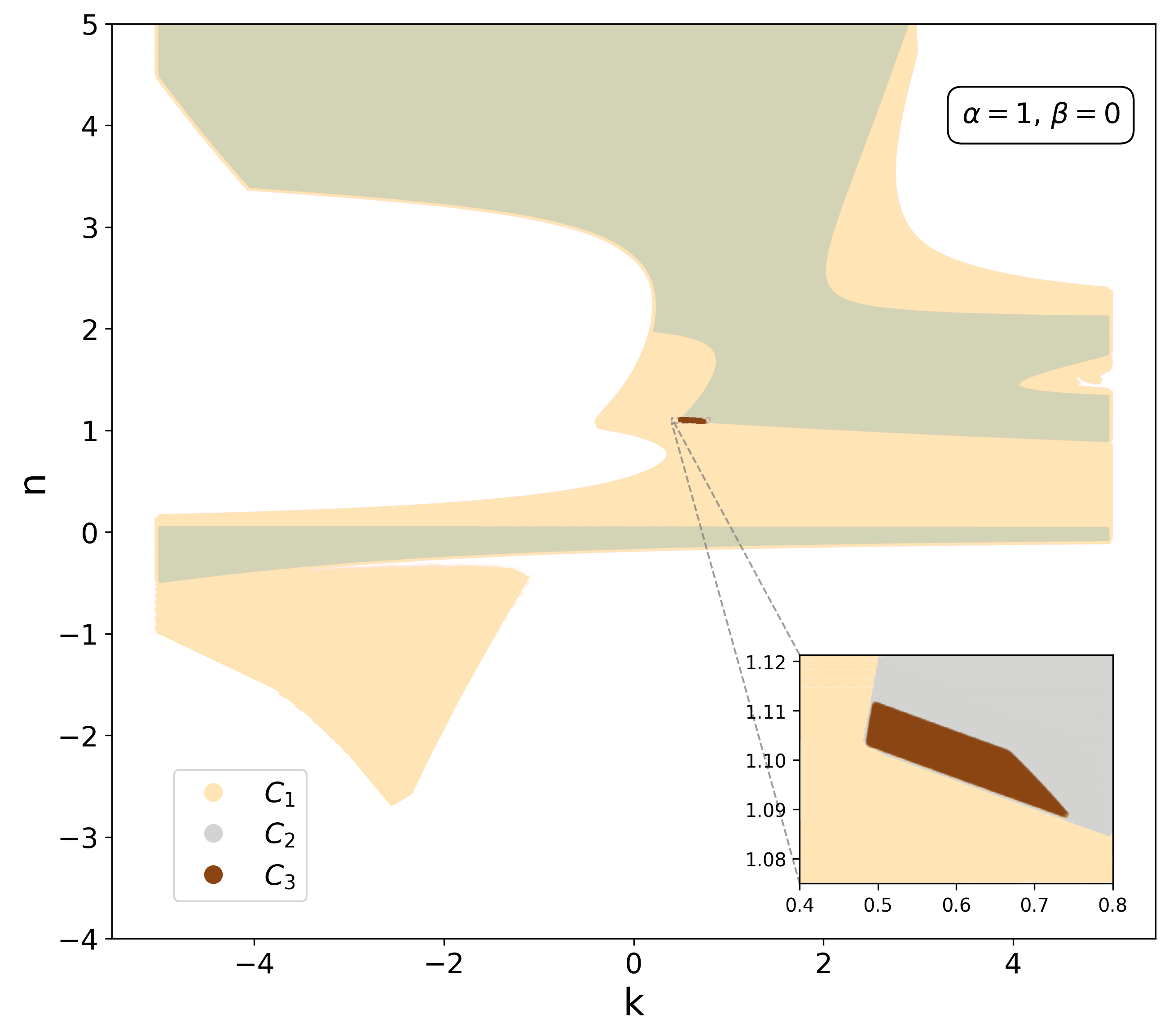} &
    \includegraphics[width=0.43\textwidth]{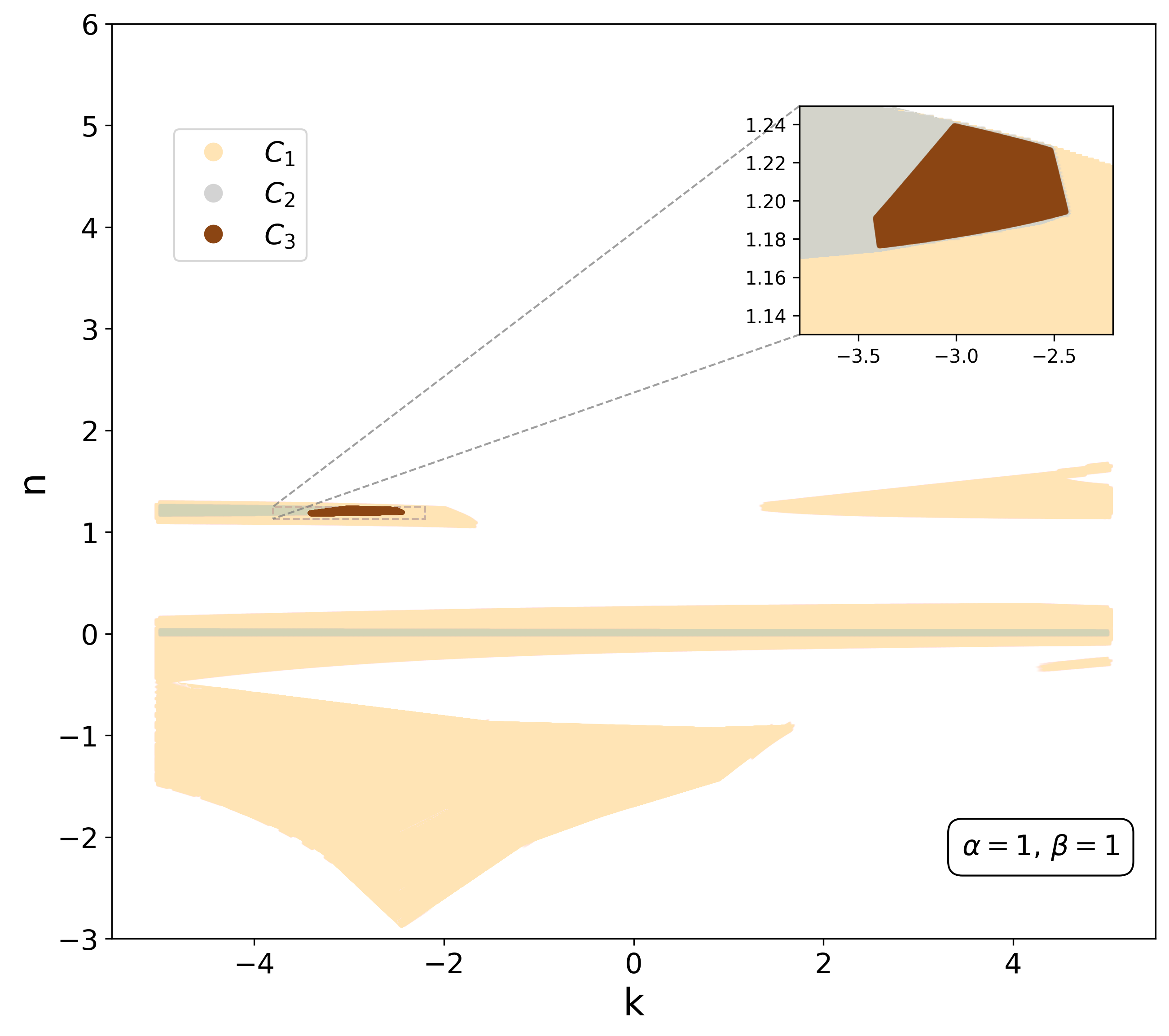} \\
    \textbf{(a)   $\alpha = 1$, $\beta = 0$} & \textbf{(b)   $\alpha = 1$, $\beta = 1$} \\[12pt]
  \end{tabular}
%  \vspace{2pt}
  
  \begin{tabular}{c}
    \includegraphics[width=0.43\textwidth]{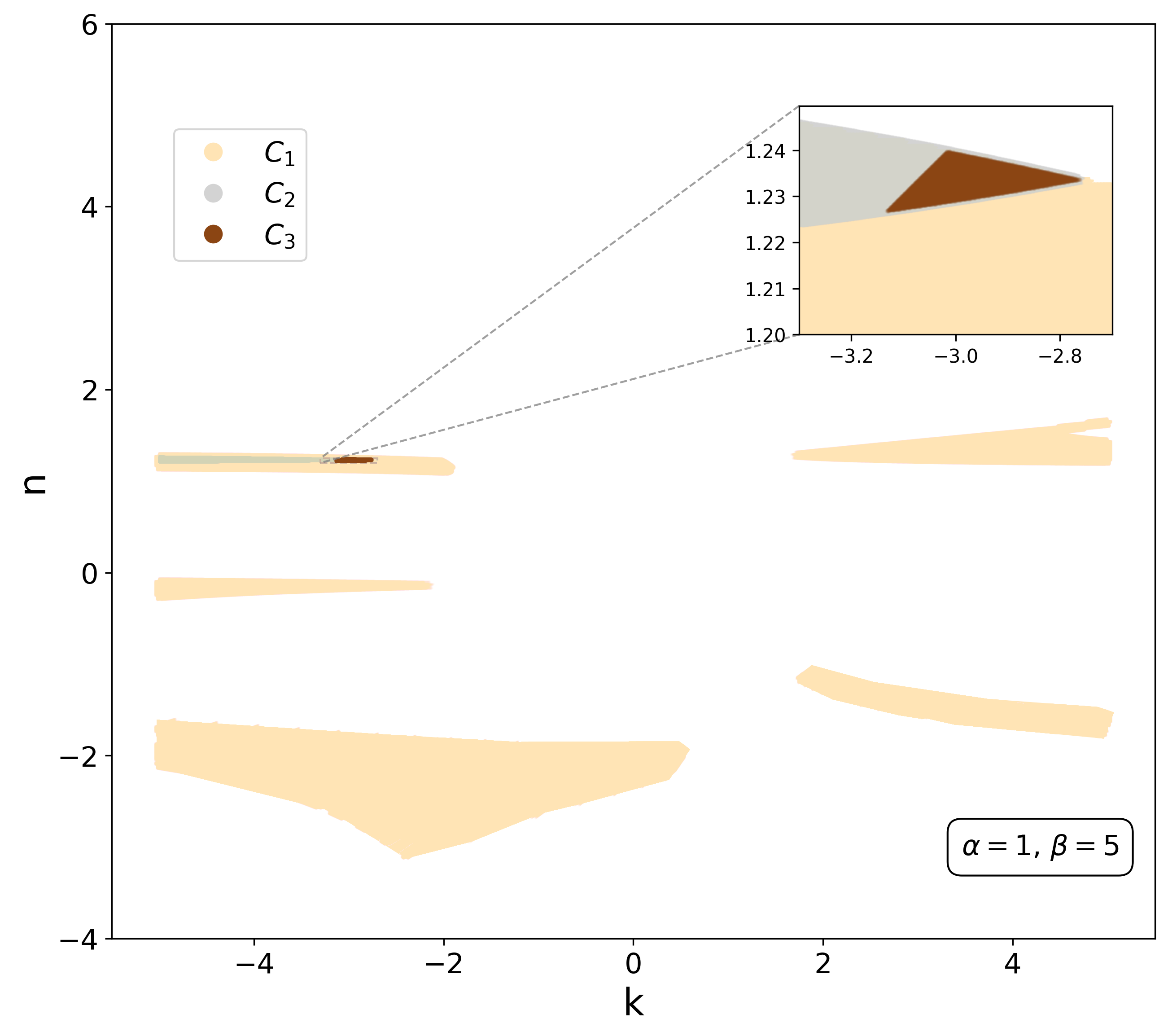} \\
    \textbf{(c)  $\alpha = 1$, $\beta = 5$} \\[6pt]
  \end{tabular}
  \caption{ Domains of the $(n-k)$ parameter space for the linear coupling scenario, for selected benchmark values of $\alpha$ and $\beta$, corresponding to the distinct highlighted regions $C_1$, $C_2$, and $C_3$ as described in the text.} 
  \label{fig:F4}
\end{figure}

In this scenario,  the difference 
$\left(\frac{\partial \mathcal{G}}{\partial \mathcal{L}_m}\mathcal{L}_m\right) - \mathcal{G}({\cal L}_m)$
represents a constant denoted as $\beta$.
This constant, together with the parameter $\alpha$, plays a crucial role in determining the coupling between the curvature and matter sectors. 
We explored the  $k-n$ parameter space
within the range
$-5 \leqslant k \leqslant 5$ and $-3\leqslant n \leqslant 5$,
identifying the distinct regions $C_1, C_2, C_3$ 
as previously described. The results for the 
 linear-coupling case are presented in three panels 
 (a), (b) and (c) of fig.\ (\ref{fig:F4}) respectively for three benchmark values:  $\beta = 0, 1, 5$,
  with $\alpha$ fixed at 1. The obtained regions are found to overlap.
  However  since the  $C_3$ region is very small in the presented scale, 
  we zoomed in on the area near $C_3$ and displayed it in a separate box for all cases. 

\subsection{Non-linear-coupling case:}
\begin{figure}[H]
  \centering
  \begin{tabular}{cc}
    \includegraphics[width=0.49\textwidth, height=2.5in]{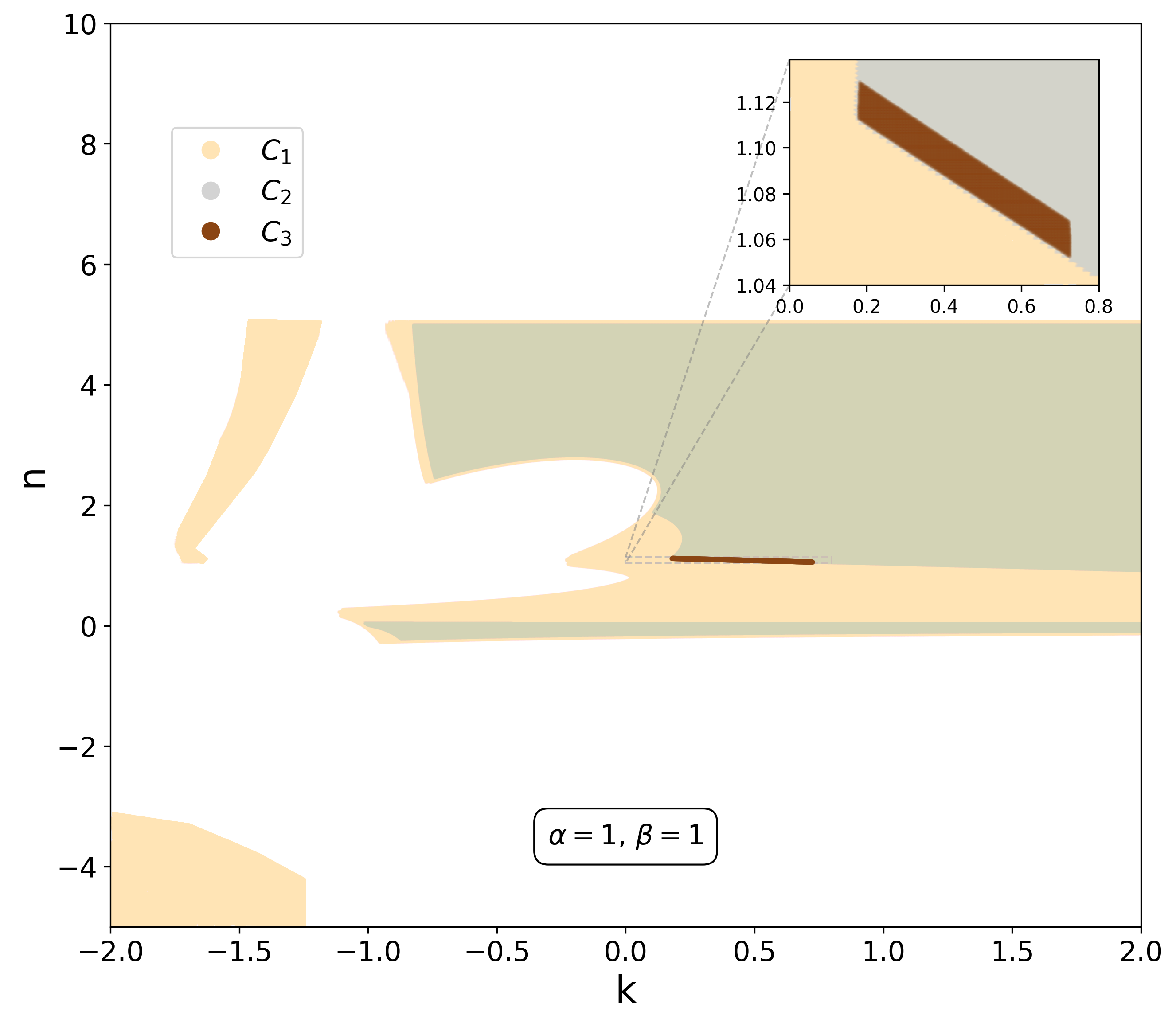} &
    \includegraphics[width=0.49\textwidth, height=2.5in]{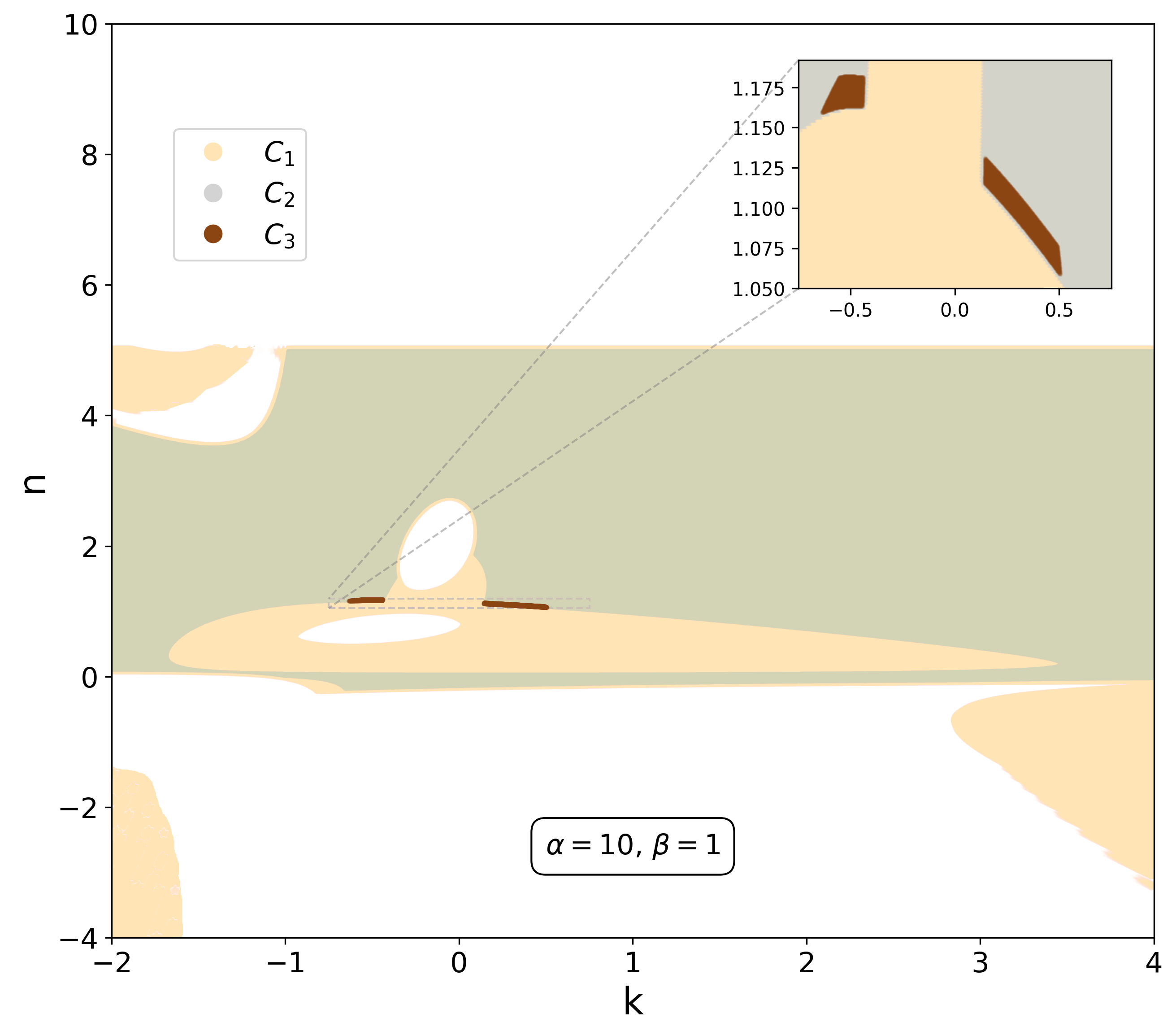} \\
    \textbf{(a)   $\alpha = 1$, $\beta = 1$} & \textbf{(b)   $\alpha = 10$, $\beta = 1$} \\[12pt]
  \end{tabular}
  
%  \vspace{2pt}
  
  \begin{tabular}{c}
    \includegraphics[width=0.49\textwidth, height=2.5in]{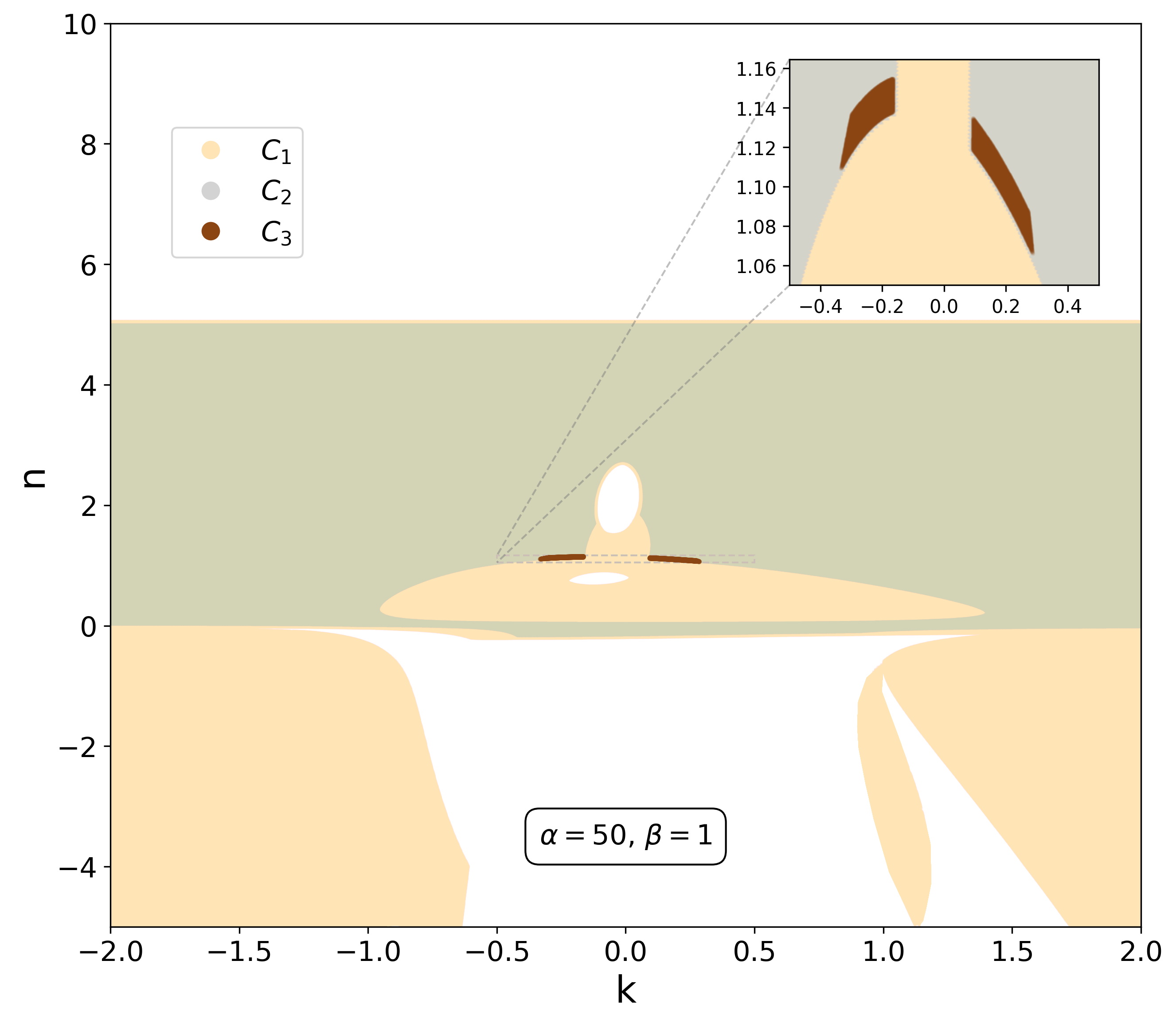} \\
    \textbf{(c)   $\alpha = 50$, $\beta = 1$} \\[6pt]
  \end{tabular}
  \caption{ Domains of the $(n-k)$ parameter space for the non-linear coupling scenario, for selected benchmark values of $\alpha$ and $\beta$, corresponding to the distinct highlighted regions $C_1$, $C_2$, and $C_3$ as described in the text.}   
   
  \label{fig:F5}
\end{figure}

In this scenario, the difference 
$\left(\frac{\partial \mathcal{G}}{\partial \mathcal{L}_m}\mathcal{L}_m\right) - \mathcal{G}({\cal L}_m)$ is considered to be proportional to $\mathcal{L}_m$ and
is denote by $\alpha\mathcal{L}_m$.
We explored the  $k-n$ parameter space
within the range  $-2\leqslant k \leqslant 2$ and $-5\leqslant n \leqslant 10$
for the non-linear coupling case.
The three regions  $C_1, C_2, C_3$ are displayed in three panels (a), (b), and (c) of
fig.\ (\ref{fig:F5}) for the three benchmark values $\alpha = 1, 10, 50$, respectively,
with $\beta$ fixed at 1 in all instances. \\

\begin{figure}[H]
  \centering
  \begin{tabular}{cc}
    \includegraphics[width=0.49\textwidth, height=2.5in]{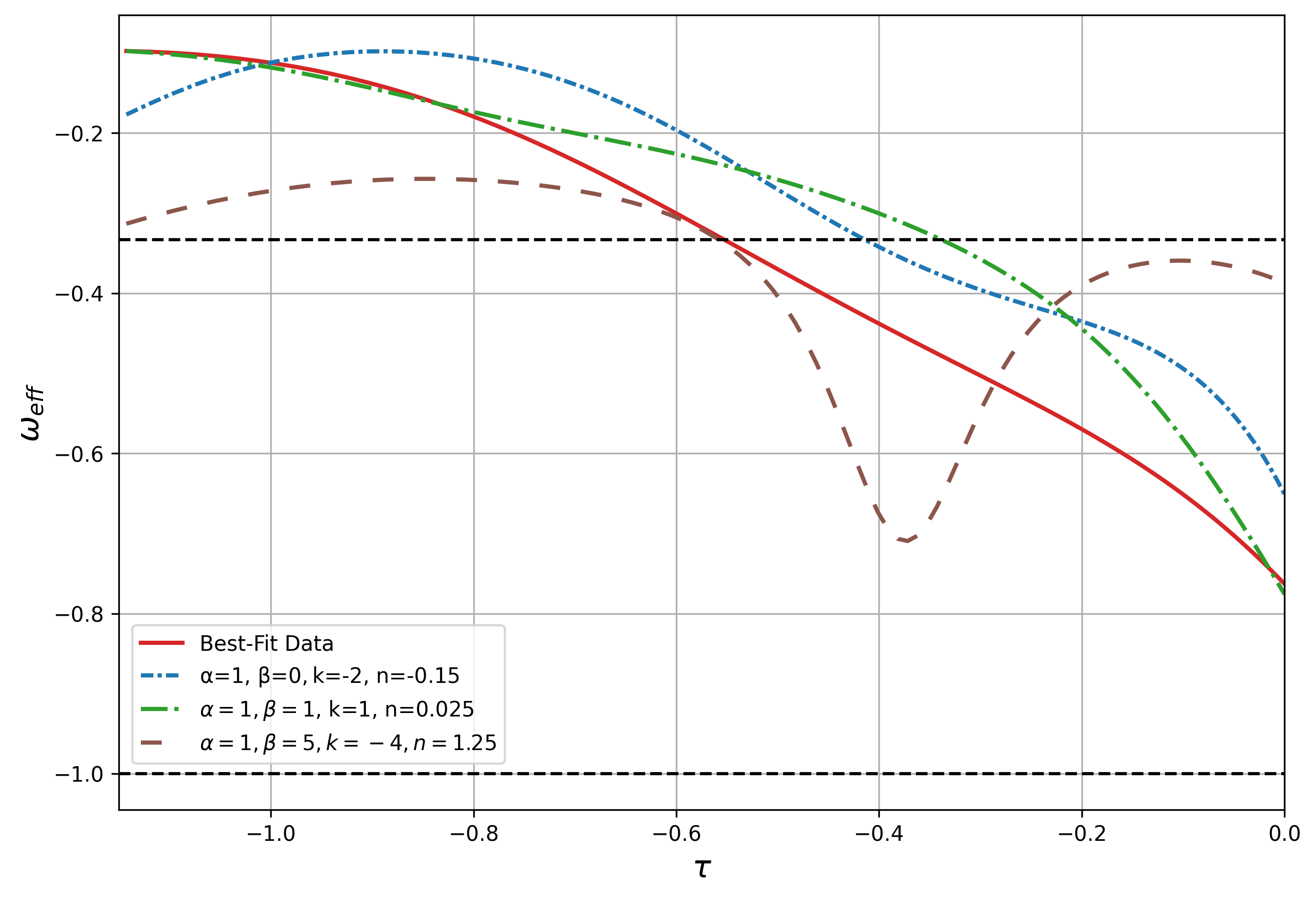} &
    \includegraphics[width=0.49\textwidth, height=2.5in]{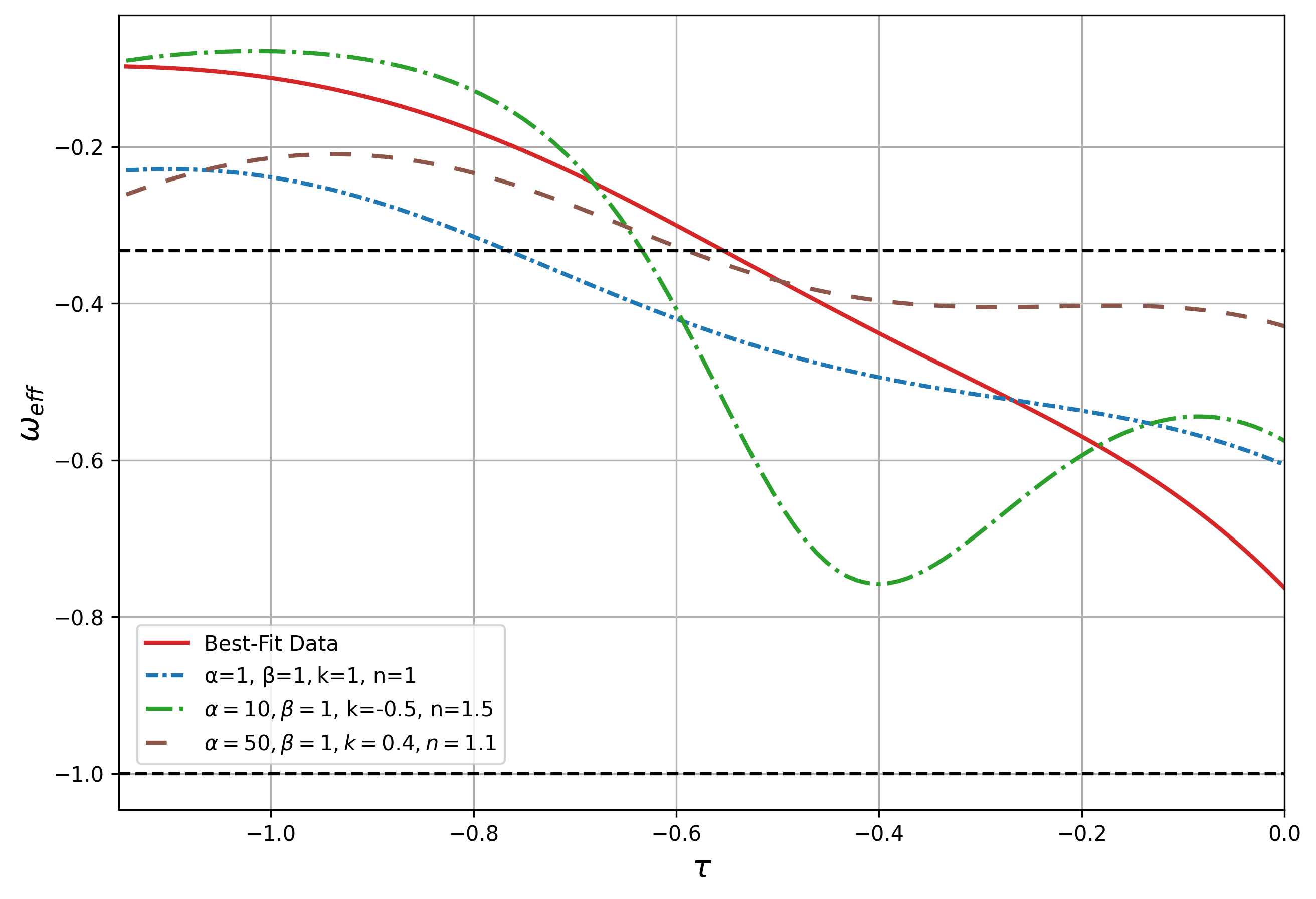} \\
    \textbf{(a) Linear-coupling case} & \textbf{(b) Non-linear coupling case} \\[12pt]
  \end{tabular}
  \caption{Temporal behavior of the effective EoS parameter $\omega_{\rm eff}$ for both linear case (left panel) and non-linear coupling case  (right panel) with benchmark values for the parameters ($\alpha, \beta, n, k$) chosen from the $C_2$ region. For comparison, the plot corresponding to the minimal coupling scenario is depicted in both panels (red curve). The horizontal dotted line represents $\omega_{\rm eff}$ value of -$\frac13$ and $-1$.}
  \label{fig:F6}
\end{figure}

Figs.\ (\ref{fig:F4}) and (\ref{fig:F5}) illustrate that both linear and non-linear coupled models exhibit specific parameter space regions ($C_1, C_2, C_3$) associated with different constraints on energy densities and pressure. This indicates that the Supernova Ia data (Pantheon), Observed Hubble data, and BAO datasets consistently support scenarios involving non-minimally coupled matter-curvature interactions. In fig.\ (\ref{fig:F6})  
we also presented the  temporal profile of the effective EoS parameter ($\omega_{\rm eff} = \frac{P}{\rho}$) for both types of models. This was done for certain 
benchmark values of the parameters  $(\alpha,\beta, n, k)$ selected from the permissible domains of the $C_2$ region for each model. \\

To express the temporal behavior of effective equation of state parameter
in fig.\ (\ref{fig:F6})
and other cosmological
quantities we use the dimensionless time parameter $\tau  = \ln\ a$.
 The features of the late-time cosmic evolution, observable in SNe Ia data from the Pantheon sample, correspond to the range $-1.18 \leq \tau \leq 0$,  where $\tau = 0$  represents the present epoch  ($a=1$). 
 The left (right) panel of fig.\ (\ref{fig:F6}) shows
 the $\omega_{\rm eff}$ vs $\tau$ plot, for  the linear (non-linear) model.
 All curves in the left (right) panel of fig.\ (\ref{fig:F6}) correspond to $\alpha=1$ ($\beta =1$). 
 In both panels, we depicted the  $\omega_{\rm eff}$ profile 
 for the minimal coupling scenario for comparison. The other curves in the left panel represent three different choices of the set $(\beta,n,k)$
(0, -2, -0.15), (1,1,0.025), (5,-4,1.25), with $\alpha$ fixed at 1.
The curves in the right panel correspond to the values of the set  
$(\alpha,n,k)$: (1,1,1), (1,-0.5,1.5), (1,0.4,1.1), with $\beta$ fixed at 1.
Both plots lie within the observationally allowed region
$C_2$  of parameter values for each model, as shown in
figs. (\ref{fig:F4}) and (\ref{fig:F5}). \\

In fig.\ (\ref{fig:F6}),  we illustrate the temporal behavior of the effective EoS parameter ($\omega_{\rm eff}$) for the minimal coupling scenario, as well as for both linear and non-linear coupling cases, for specific benchmark values for the parameters ($\alpha, \beta, n, k$) chosen from the $C_2$ region, which corresponds to the range $-1 \leqslant \omega_{\rm eff} \leqslant 0$. The plots for both minimal and non-minimal coupling scenarios demonstrate a transition in the $\omega_{\rm eff}$ value from above to below $-1/3$ during the late time phase  of cosmic evolution. In the linear coupling case, the $\omega_{\rm eff}$ profile for lower values of the coupling parameter $\beta$ closely aligns with that of the non-minimal coupling scenario. However, for a higher value of $\beta \sim 5$, there is a noticeable shift, exhibiting a bouncing behavior in the effective EoS parameter profile as evident from the plot. In the non-linear coupling case, for lower and moderate values of the coupling parameter $\alpha$, the EoS parameter stabilizes around -0.6, similar to that obtained for the non-minimal coupling scenario. In both linear and non-linear coupling scenarios, a non-phantom accelerating dark energy solution is observed at the selected benchmark values of the non-minimally coupled model.\\

\begin{figure}[H]
\centering
\includegraphics[width=0.7\textwidth, height=3in]{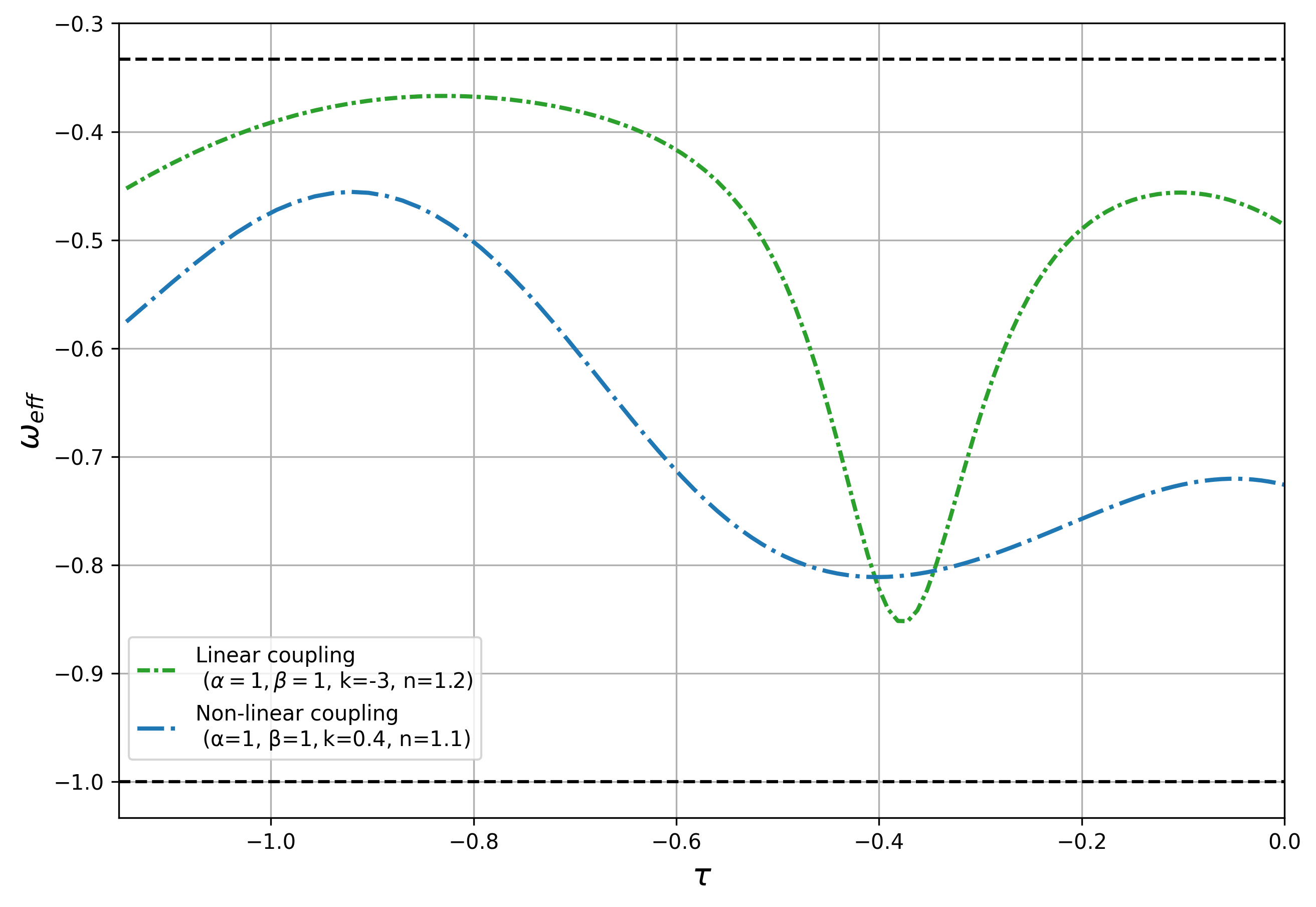}
\caption{Comparison between temporal behavior of accelerated effective EoS parameter ($\omega_{\rm eff}$) for both linear and non-linear coupling scenarios with parameter values chosen from $C_3$ region.} 
\label{fig:F7}
\end{figure}

To further analyze the features of temporal variations in the effective EoS parameter, we investigate the variation of the profile for benchmark values of model parameters $(k,n)$ chosen from the $C_3$ region  (with coupling
parameters $\alpha$ and $\beta$ fixed at 1). The corresponding $\omega_{\rm eff}$ profiles for both linear and nonlinear coupling cases are presented in fig.\  (\ref{fig:F7}),  indicate that the profiles are sensitive to the parameters $n$ and $k$. The profile for the linear coupling scenario demonstrates a non-phantom type of dark energy behavior throughout the accessible time domain. Near $\tau = -0.4$, there is a sharp dip in the EoS parameter, bringing its value close to $-1$. As time progresses, the value of the EoS parameter increases to $-0.5$ and stabilizes there. A similar non-phantom dark energy behavior is observed in the nonlinear coupling case also, where the EoS parameter finally settles near $-0.7$, implying that the non-linearly coupled curvature-matter model too exhibits non-phantom type dark energy at the chosen benchmark values (as mentioned in the caption of  fig.\ (\ref{fig:F7})).

\begin{figure}[H]
\centering
\includegraphics[width=0.7\textwidth, height=3in]{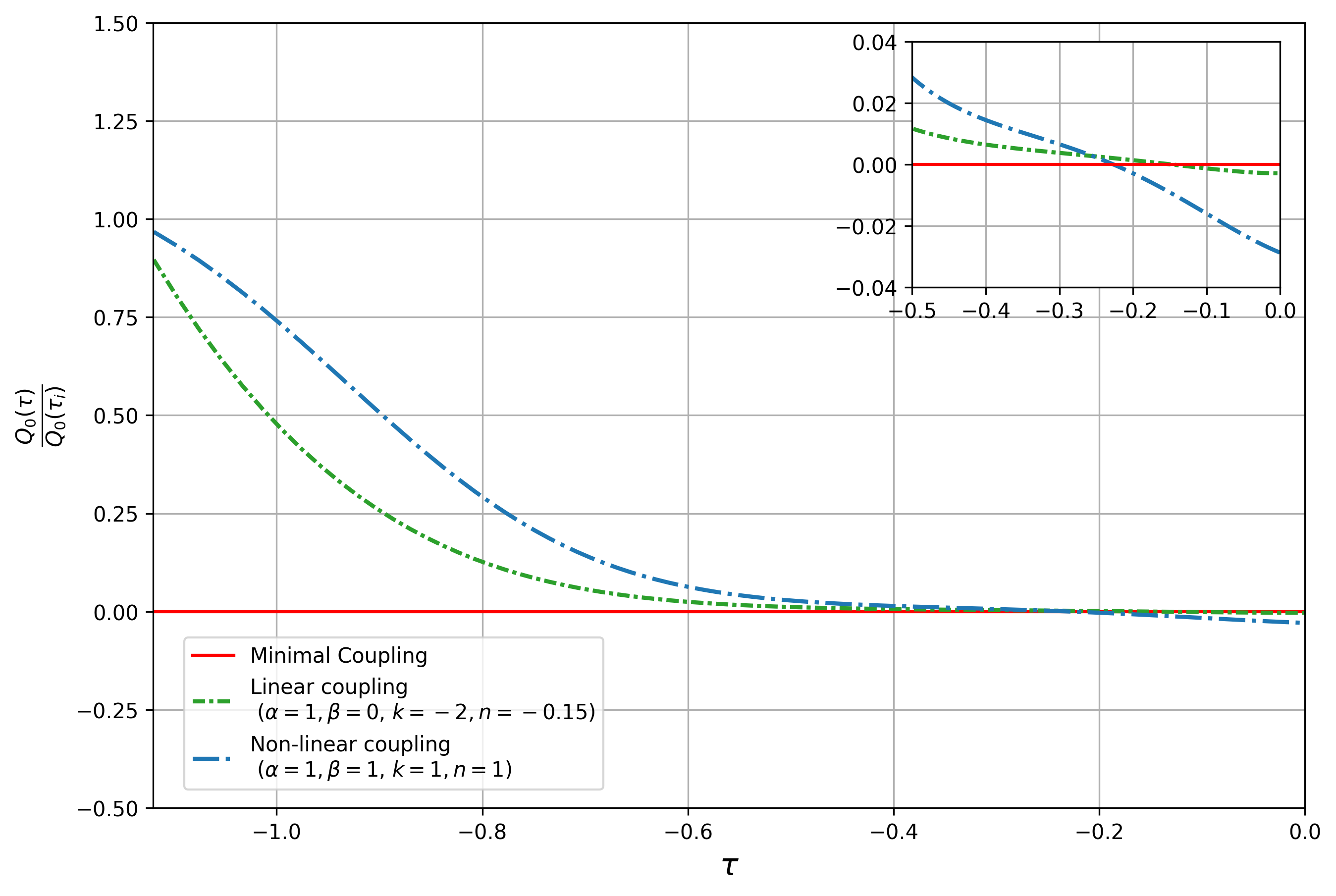}
\caption{Temporal behaviour the 0-th component of
the normalized source term 
 (discussed in text) at two different coupling cases 
for parameter values  $(\alpha,\beta, n, k)$ chosen from $C_2$ region.} 
\label{fig:F8}
\end{figure}

We investigate the time dependence of the term source term
in the  non-conserving continuity equation for the matter 
and curvature sectors
arising from specific choices of the matter Lagrangian within this coupled theory.   
As evident from the eq.\ (\ref{eq:b4}), the choice of matter lagrangian
$\mathcal{L}_m = p$ ensure presence of non-zero source term in the continuity 
equation for both types of coupling and   its '0' component,    specified in 
 eqs.\ (\ref{eq:b12}) and (\ref{eq:c3})
  for linear and non-linear coupling scenarios, respectively
  and it represents the rate of energy transfer between these two sectors. 
$Q_0$  depends on the coupling and model parameters and for 
a given set of parameter values, the temporal behavior of $Q_0$ 
can be computed using the observed time dependences of the scale 
factor and Ricci scalars, as well as the time dependences of the energy 
density and pressure for both types of coupling scenarios.
To illustrate the variation of $Q_0$ 
with respect to the time parameter $\tau$
we have plotted the dimensionless quantity $Q_0/Q_0(\tau_{\rm in})$
in fig.\ (\ref{fig:F8}),
where  $\tau_{\rm in}$  denotes the initial instant corresponding to the earliest epoch accessible in the observational data considered.
We have shown these plots for both  linear and nonlinear coupling scenarios. \\
  
 The magnitude of the source term is observed to be higher in the early universe. As time progresses, its absolute value decreases rapidly for both types of coupling scenarios. 
Around $\tau \sim -0.4$, the source term approaches zero, indicating the dissipation of non-conserving effects. A closer examination of the plot (highlighted in the zoomed portion of  
  fig.\ (\ref{fig:F8}), reveals a sign flip in the source term for both scenarios around the epoch marked by $\tau \sim -0.2$.
 However, around this epoch, the value of the source term is negligibly small compared to that of earlier epochs.\\

The observed dynamics of energy exchange, as indicated by the temporal variation of $Q_0$ for chosen parameter values $(\alpha, \beta, n, k)$ from the region $C_2$, suggest that the integration of Pantheon, OHD, and BAO data permits non-minimal time-varying curvature and matter interactions, leading to energy exchange between the two sectors. This dynamics can be expressed by the energy balance equation in the FLRW background $\dot{\rho} + 3H(\rho + p + p_c) = 0$, where  $p_c$
encapsulates the effect of the non-minimal coupling between curvature and matter. From a thermodynamic perspective of open systems, as extensively discussed in \cite{Harko:2015pma}, curvature-matter coupling facilitates the production of a substantial amount of comoving entropy during the late-time evolutionary phase of the universe. This enables the interpretation of the energy-balance equation in terms of particle creation in the FRW universe, where $p_c$ is identified as the  creation  pressure associated with particle creation. This interpretation, however, necessitates  $p_c$ to be negative, which is consistent with the chosen parameter values within the range  $C_2$. Therefore, the `particle creation' interpretation of the energy balance equation in non-minimally coupled curvature-matter scenarios remains a viable possibility in the context of the combined Pantheon, OHD, and BAO data.

%%%%%%%%%%%%%%%%%%%%%%%%%%%%%%%
%%%%%%%%%%%%%%%%%%%%%%%%%%%%%%% 
\section{Conclusion}
\label{Sec:V}
  In this article we investigate the  gravity models with non-minimal curvature-matter coupling and their cosmological implications within a flat-FLRW spacetime framework, using observational data. The non-minimal coupling is introduced via a modified Einstein-Hilbert action, incorporating the term $ 
\int \sqrt{-g} \, d^4x \, \mathcal{G}(\mathcal{L}_m) f_2(R)$.
For specific forms of $\mathcal{G}(\mathcal{L}_m)$ and $f_2(R)$, the action reduces to the minimally coupled Einstein-Hilbert action.
Key parameters include $\alpha$ and $\beta$ (strength of non-minimal coupling), $n$ (parametrizing $f_2(R)$), and $k$ (parametrizing pressure $p$ in the matter Lagrangian $\mathcal{L}_m = p$). These parameters influence the interaction dynamics, which are examined assuming a homogeneous, isotropic flat FLRW metric with scale factor $a(t)$.
The matter content is modeled as a perfect fluid with energy density $\rho(t)$ and pressure $p(t)$. The choice $\mathcal{L}_m = p$ ensures hydrodynamical consistency and addresses deviations from geodesic motion in non-minimally coupled scenarios. This leads to a non-conserving continuity equation, facilitating energy exchange between matter and curvature sectors.
Field equations derived from the modified action connect $a(t)$, $\rho(t)$, $p(t)$, their time derivatives, and functions like the Hubble parameter $H(t)$ and Ricci scalar $R(t)$. These equations depend on $\alpha$, $\beta$, $n$, and $k$, and reduce to standard Friedmann equations when $\beta = 0$ and $\alpha = 1$.  Unlike the case for Friedman equations, 
the modified field equations   involve time derivatives of $p$.
Observational constraints are applied to an exponential pressure model, $p(t) = p_0 \exp(ak)$, to assess the viability of non-minimally coupled fluid-curvature models with the four parameters $(\alpha, \beta, n, k)$. \\

  Through a comprehensive analysis of the Pantheon compilation of SNe Ia data, combined with OHD and BAO datasets, we derived the time evolution of key cosmological quantities, including $a(t)$, $H(t)$, and $R(t)$, along with their time derivatives, during the later stages of cosmic evolution. Using these datasets and the evolution equations for non-minimally coupled models, we numerically computed the energy density $\rho(t, \alpha, \beta, n, k)$, pressure $p(t, \alpha, \beta, n, k)$, and the effective equation of state (EoS) parameter profiles for linear and non-linear coupling scenarios with an exponential pressure profile.
Domains in the $k$-$n$ parameter space were identified where observational data supports the specific energy conditions: (i) Non-negative energy density satisfying the Weak Energy Condition  (ii) the Null, Weak, and Dominant Energy Conditions are simultaneously satisfied
(iii) a non-phantom dark energy era satisfying satisfying null and Weak energy conditions but violating strong energy conditions.  
The temporal behavior of the effective EoS parameter for these cases was presented for both coupling scenarios. In both models, we observed an accelerating, non-phantom dark energy solution consistent with observational data. Both types of non-minimal coupling can partially replicate the behavior of the minimally coupled scenario. \\

We computed the temporal profile of the source term in the continuity equation within the framework of the coupled theory, considering dependencies on higher-order Ricci curvature terms and different coupling types.   For the linear coupling case, the source term depends on pressure, energy density, the Ricci scalar, its time derivative, coupling strength (\(\alpha\)), and curvature parameter (\(n\)). In the non-linear case, the dependence becomes more complex. Fig. (\ref{fig:F8}) shows the normalized energy transfer function over time for both cases, differing from the conserving case (red line). A non-zero source term enables particle creation, governed by the creation pressure (\(p_c\)), which becomes negative if the effective EoS parameter lies between 0 and -1. The figure also highlights a crossover point where, for certain parameters, the source term vanishes, mimicking the minimally coupled scenario. At a specific moment, the energy transfer halts before resuming, alternating between the two sectors. 
The absolute magnitude of the source term, which measures the rate of energy exchange between the curvature and matter sectors, is found to decrease over time, reaching zero around the epoch marked by $\tau = -0.22$. After this epoch, a slight deviation from the zero value of the source term is observed. The sign of the source term flips after crossing $\tau = -0.22$, indicating a reversal in the direction of energy exchange before and after this instant. This has been
observed for both linear and non-linear coupling
scenarios for paramter-set $(\alpha, \beta, n, k)$ values chosen from the region $C_2$. From a thermodynamic perspective,  this energy exchange dynamics can be interpreted  in terms of particle creation in the FLRW universe,   influenced by a particle creation pressure $p_c$ that appears in the energy balance equation   $\dot{p} + 3H(p + \rho+ p_c) = 0$ which describes the dynamics of energy exchange. \\

Thus, our investigation highlights the significant potential of non-minimally coupled curvature-matter models in explaining the observed accelerated expansion of the universe. By incorporating observational data from Pantheon, BAO and OHD, we have shown that these models can provide a consistent description of the  late-time evolutionary dynamics of the universe. The inclusion of non-minimal coupling parameters $\alpha$ and $\beta$ along with model parameters $n$ and $k$ offers a rich framework for exploring various cosmological scenarios, including those that deviate from standard cosmology.   One of the limitations of this non-minimally coupled curvature-matter model is its inability to represent the evolution of all phases of the universe (radiation, matter, and dark energy), as it primarily focuses on depicting a non-phantom dark energy-dominated scenario. To address this, future studies could explore modified gravity models like \( f(R) \) or \( f(R, T) \), which encompass all phases of cosmic evolution. Using a relativistic fluid model instead of a specific fluid model may also enhance the framework. Further investigations could include the growth of matter perturbations and gravitational wave generation in the non-minimal coupling context. 
Additionally, we adopt a non-linear form for \( {\cal G}({\cal L}_m) \) such that the modified field equations depend only on \( {\cal G}({\cal L}_m) \), avoiding derivative terms with respect to \( {\cal L}_m \). This simplifies the analysis, focusing on non-minimality effects from non-linearity in \( {\cal G}({\cal L}_m) \). While a full analysis of alternative non-linear forms with appropriate parameters remains an open task, it is beyond the scope of this work.  \\

Our results indicate that non-minimal couplings can mimic the effects of dark energy without invoking exotic fields, suggesting a more nuanced interaction between matter and curvature. Future work in this line will involve refining these models with more precise data and extending the analysis to include other cosmological phenomena, potentially offering deeper insights into the fundamental nature of gravity and the expansion of the universe.

\paragraph{Acknowledgement}\
All authors would like to thank the referees for their valuable suggestions.

\end{document}